\documentclass[12pt]{iopart}

\usepackage{siunitx}
\usepackage{subfig}
\usepackage{enumerate}
\usepackage{xcolor}
\usepackage{titlesec}
\usepackage{graphicx}
\usepackage{verbatim}
\usepackage{booktabs}
\usepackage{caption}

\usepackage{rotating}
\usepackage{comment}

\usepackage{cite}

\begin{document}
	
	\title[Total nuclear reaction cross-section database]{Total nuclear reaction cross-section database for radiation protection in space and heavy-ion therapy applications}
	
	\author{F Luoni$^{1,2}$, F Horst$^1$, CA Reidel$^1$, A Quarz$^{1,3}$, L Bagnale$^1$, L Sihver$^{4,5}$, U Weber$^1$, RB Norman$^6$, W de Wet$^7$,  M Giraudo$^8$, G Santin$^{9,10}$, JW Norbury$^{6}$, M Durante $^{1,2}$}
	
	\address{$^1$ Biophysics Department, GSI Helmholtzzentrum für Schwerionenforschung GmbH, Darmstadt, Germany \\$^2$ Technische Universität Darmstadt, Institut für Physik Kondensierter Materie, Darmstadt, Germany \\$^3$ Hochschule Darmstadt, Institut für Fachbereich Matematik und Naturwissenschaft \& Informatik, Darmstadt, Germany \\$^4$ Technische Universität Wien, Atominstitut, Vienna, Austria \\$^5$ Chalmers University of Technology, Gothenburg, Sweden \\$^6$ NASA Langley Research Center, Hampton, VA, USA\\ $^7$ University of New Hampshire, Durham, NH, USA \\$^8$ Thales Alenia Space Italia, Torino, Italy \\$^9$ ESA ESTEC, Noordwijk, Netherlands \\$^{10}$ RHEA System, Noordwijk, Netherlands\\}
	\ead{m.durante@gsi.de}
	\vspace{10pt}
	\begin{indented}
		\item[]
	\end{indented}
	
	\begin{abstract}
		Realistic nuclear reaction cross-section models are an essential ingredient of reliable heavy-ion transport codes. Such codes are used for risk evaluation of manned space exploration missions as well as for ion-beam therapy dose calculations and treatment planning. Therefore, in this study, a collection of total nuclear reaction cross-section data has been generated within a GSI-ESA-NASA collaboration. The database includes the experimentally measured total nucleus-nucleus reaction cross-sections.
		The Tripathi, Kox, Shen, Kox-Shen, and Hybrid-Kurotama models are systematically compared with the collected data. Details about the implementation of the models are given.
		Literature gaps are pointed out and considerations are made about which models fit best the existing data for the most relevant systems to radiation protection in space and heavy-ion therapy.
	\end{abstract}
	
	%
	\vspace{2pc}

	\section{Introduction}
	
	The dangers due to galactic cosmic rays (GCR) are the biggest hindrance to manned long-term deep-space exploration missions \cite{Durante2011,Chancellor2014}. The galactic and extra-galactic cosmic ray spectrum consists of all natural elements and their kinetic energies range from about $\SI{10}{MeV/u}$ to the $\si{ZeV}$ region. Elements heavier than nickel are present only in trace amounts. 
	As GCR are a very difficult radiation environment to be reproduced on Earth, stochastic (Monte Carlo) or deterministic transport codes are needed for risk assessment of exploration mission scenarios and shielding design.
	Similar transport codes are also used in ion-beam therapy dose calculations and treatment planning. Heavy-ions such as $^{4}$He or $^{12}$C are exploited for radiation therapy because of their favorable depth-dose profile (Bragg curve). $^3$He and $^{16}$O were also proposed to be used for heavy-ion therapy purposes \cite{Horst2021,Sokol2017}. The shape of the Bragg peak, the tail behind it and the entrance channel of the curve differ for every ion because of the nuclear interactions the ion undergoes when travelling through the human body \cite{Haettner2013,Schardt2010}.
	Realistic models for such cross-sections are an essential ingredient to reliable deterministic and stochastic transport codes \cite{Townsend2002,Norbury2012}. In Monte Carlo (MC) codes the nuclear interaction distance of heavy-ions with matter is sampled from a probability function that depends on total reaction cross-sections $\sigma_\textup{R}$.
	Examples of how much the choice of cross-section parameterization can influence dose calculations in heavy-ion therapy can be found in Ref.~\cite{Arico2019,Luehr2012}.
	
	Currently, there is no total nuclear reaction cross-section formula that fits well the experimental data for all projectile-target systems. Several semi-empirical parameterizations have been proposed in the last decades and some of them are implemented in the MC and deterministic tools most commonly used for space radiation protection and heavy-ion therapy applications, such as the MC codes Geant4 \cite{G4_cern,Agostinelli2003,Allison2006,Allison2016}, PHITS \cite{PHITS}, FLUKA \cite{FLUKA_link,Andersen2004} and the deterministic HZETRN \cite{HZETRN}. The parameterizations were often compared only to a limited data set \cite{Sihver2012,Boehlen2010}.
	
	The aim of the present work is to give a broad overview of all nucleus-nucleus total reaction cross-section data measured so far and to compare them with the most-commonly used parameterizations in transport codes. For this purpose, a total reaction cross-section data collection has been created within a collaboration between GSI Helmholtzzentrum für Schwerionenforschung (Darmstadt), the European Space Agency ESA and the National Aeronautics and Space Administration NASA, in the framework of the ROSSINI3 project. 
	
	The link to the web application where an up-to-date version of the data collection and a tool to generate cross-section plots as a function of kinetic energy is: \texttt{https://www.gsi.de/fragmentation}. Details about the data collection, the implemented parameterizations and the web application are given in the sections~\ref{sec:database},~\ref{sec:formulae} and~\ref{sec:webpage}.
	Through such a comprehensive study, recommendations are given of what cross-section data should be measured in future experiments and what formulae fit the existing data best for the most relevant systems to radiation protection in space and heavy-ion therapy.
	
	\section{Data collection description}
	\label{sec:database} 
	
	A comprehensive data collection was generated by Norbury \textit{et al.} \cite{Norbury2012}. It analyzed measured cross-section data relevant to radiation protection in space. The information concerned the systems of target and projectile for which the cross-section data were measured, the type of cross-section 
	and the kinetic energy range of the projectile. However, the exact energy of the projectile and the measured cross-section values were not reported in the review. 
	
	The data collection presented in the present work is the result of a collaboration between the GSI and the space agencies ESA and NASA. It has been decided to focus on nucleus-nucleus reactions. For this reason, reaction cross-section data measured for hadronic projectiles such as protons and neutrons are not reported. Only English peer-reviewed works have been included.
	
	A total of 1786 cross-section data from 103 publications \cite{Horst2021,Horst2019,Zhang2012,Hirzebruch1995,Hirzebruch1993,Ozawa2014,Cecchini2008,Matsuoka1980,Giot2013,Paradela2017,AlcantaraNunez2015,Ferrando1988,Horst2017,Li2016,Gupta2013_1,Alpat2013,Ozawa1996,Ozawa1995,Ozawa1994,Westfall1979,Shapira1982,Cai2002,Fang2001,Budzanowski1967,Ingemarsson2000,Labie1973,Togo2007,Cheng2012,Bisheva1967,Tran2016,Yamaguchi2011,Tanihata1992,Tanihata1985_1,Tanihata1985_2,Tanihata1985_3,Tanihata1988,Kobayashi1992,Bochkarev1998,Gupta2012,Gupta2013_2,Greiner1985,Wang2019,Napolitani2007,Webber1982,Binns1987,Brechtmann1988_1,Brechtmann1988_2,Brechtmann1988_3,Brechtmann1989,Brohm1995,Chen1994,Cheshire1974,Christie1993,Chulkov2000,Flesh2001,Golovchenko2001,Golovchenko2002,He1994,Hirzebruch1992,Iancu2005,Nilsen1995,Price1991,Sampsonidis1995,Schall1996,Toshito2007,Webber1990,Webber1998_1,Webber1998_2,Yamaguchi2010,Zeitlin2001,Zeitlin2007_1,Zeitlin2007_2,Zeitlin2008,Zeitlin2011,Goekmen1984,Igo1963,DeVries1982,Millburn1954,Bilaniuk1981,Perrin1982,Kox1984,Takechi2009,Singh1992,Aksinenko1980,Hostachy1987,Fukuda1991,Fukuda1999,Kox1987,SaintLaurent1989,Suzuki1995,Suzuki1998,Obuti1996,Blank1997,Fang2000,Ozawa2001,Powers1966,Jaros1978,Neumaier2002,Warner1996,Ingemarsson2001,Auce1996,Wilkins1962,Mayo1965} have been included in the database so far. If old data were replaced, only the newer dataset has been included (e.g. data from Ref.~\cite{Auce1994} have not been reported as they were replaced with data from Ref.~\cite{Ingemarsson2000}).
	
\begin{figure}[htb]
		\centering
		\subfloat[][All entries]
		{\includegraphics[width=.65\textwidth]{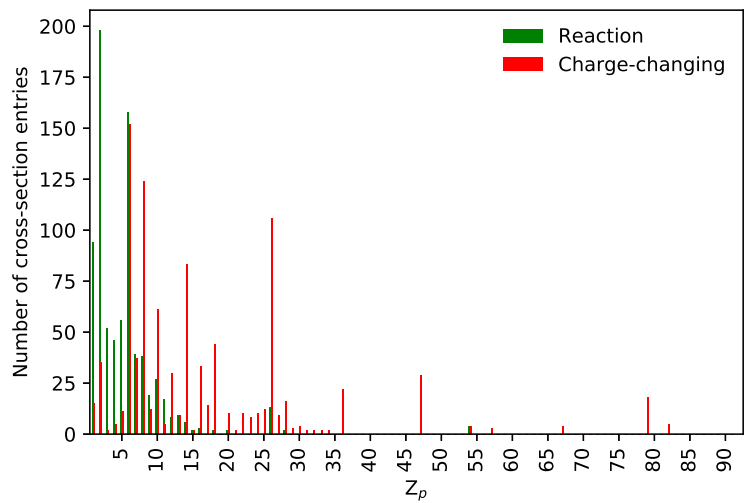}} \quad
		\subfloat[][Entries up to Ni projectiles]
		{\includegraphics[width=.65\textwidth]{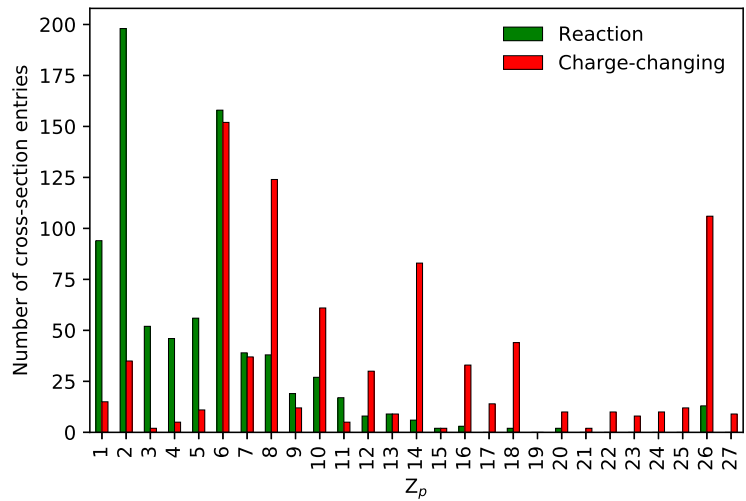}}
		\caption{Number of cross-section data reported in the database as a function of the atomic number of the projectile nucleus $Z_p$. Charge-changing cross-section entry numbers are shown in red and reaction in green. In panel (a) all entries are shown, in panel (b) only entries up to nickel projectiles.}
		\label{fig:counts}
	\end{figure}

	In Figure~\ref{fig:counts}, the number of cross-section data reported in the database are plotted as a function of the atomic number of the projectile nucleus. A zoom on data points up to Ni ions is also depicted, since nuclei heavier than Ni are only present in trace amounts in the GCR spectrum. The contribution of charge-changing cross-sections (detailed in section~\ref{sec:cc_mc}) is shown in red and of reaction cross-sections in green. 
	
\begin{figure}[htb]
		\centering
		\subfloat[][]
		{\includegraphics[width=1\textwidth]{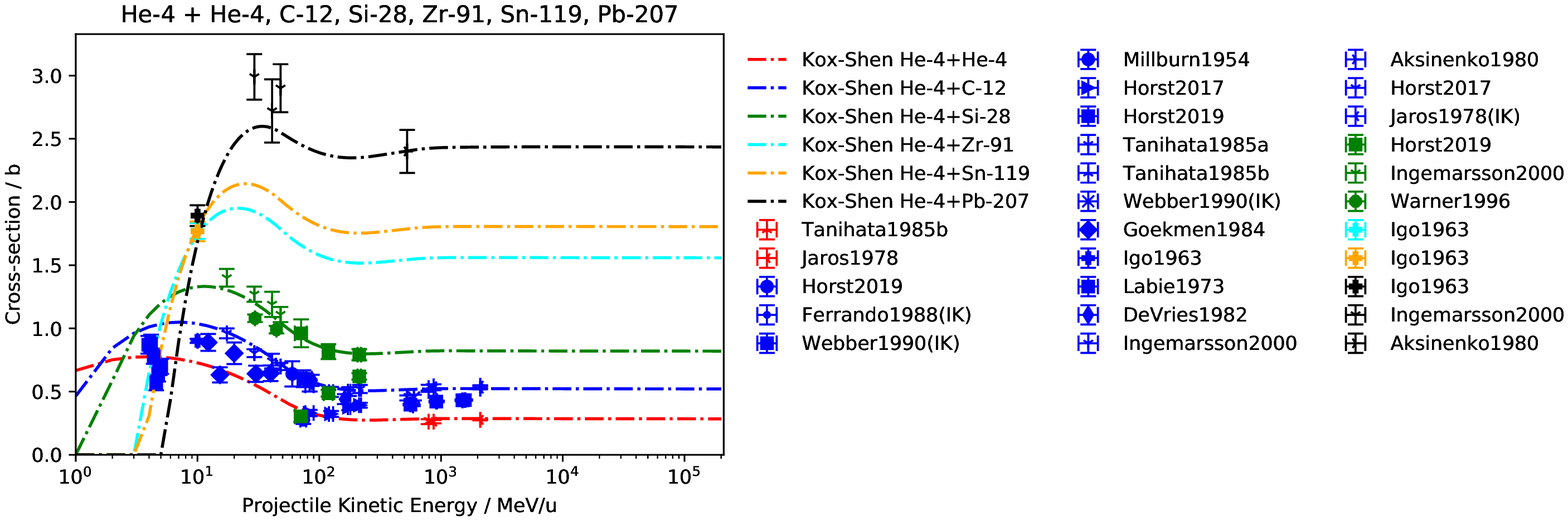}}\\
		\subfloat[][]
		{\includegraphics[width=1\textwidth]{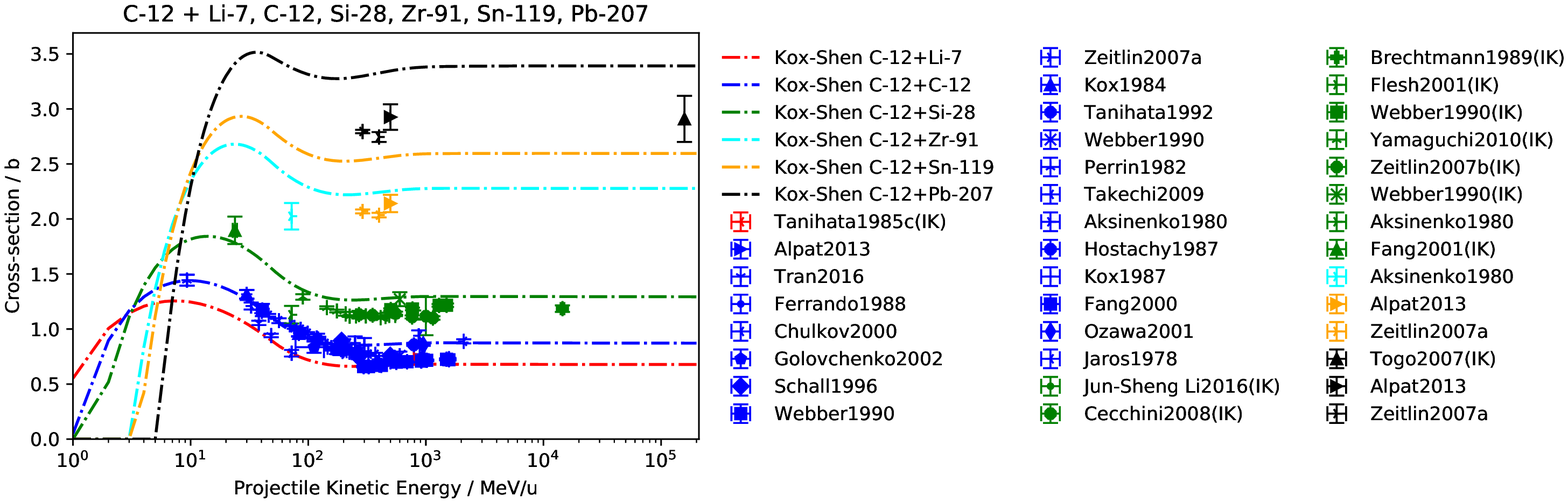}}\\
		\subfloat[][]
		{\includegraphics[width=1\textwidth]{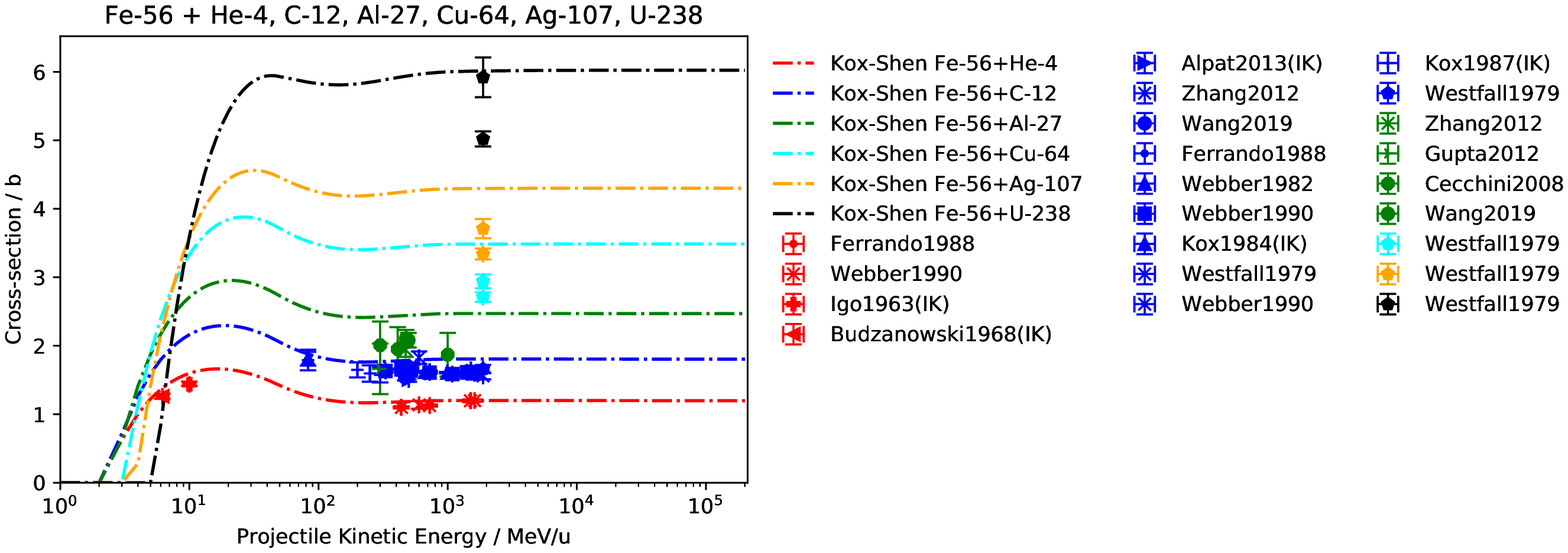}}
		\caption{Data collections for $^4$He, $^{12}$C and $^{56}$Fe projectiles on different targets with the predictions of the Kox-Shen semi-empirical model (see section~\ref{sec:formulae}) to guide the reader's eyes. IK stands for Inverse Kinematic data (see section~\ref{sec:formulae}). Different colors represent different targets. Both reaction and charge-changing cross-sections are plotted. For $^{56}$Fe + $^{64}$Cu, $^{107}$Ar, $^{238}$U (panel~(c)) both charge- and mass-changing cross-section data were measured by Westfall \textit{et al.} \cite{Westfall1979}. Therefore, two data points are reported for each system for the same energy. 
		Data in panel (a) are from references~\cite{Tanihata1985_1,Tanihata1985_2,Jaros1978,Horst2019,Ferrando1988,Webber1990,Millburn1954,Horst2017,Goekmen1984,Igo1963,Labie1973,DeVries1982,Ingemarsson2000,Aksinenko1980,Warner1996}. Data in panel (b) are from references~\cite{Tanihata1985_3,Alpat2013,Tran2016,Ferrando1988,Chulkov2000,Golovchenko2002,Schall1996,Webber1990,Zeitlin2007_1,Kox1984,Tanihata1992,Webber1990,Perrin1982,Takechi2009,Aksinenko1980,Hostachy1987,Kox1987,Fang2000,Ozawa2001,Jaros1978,Li2016,Cecchini2008,Brechtmann1989,Flesh2001,Yamaguchi2010,Zeitlin2007_2,Togo2007}.
		Data in panel (c) are from references~\cite{Ferrando1988,Webber1990,Igo1963,Budzanowski1967,Alpat2013,Wang2019,Zhang2012,Webber1982,Webber1990,Kox1984,Westfall1979,Kox1987,Gupta2012,Cecchini2008}.  
		}
		\label{fig:data}
	\end{figure}

	Data collections for $^4$He, $^{12}$C and $^{56}$Fe projectiles impinging on different targets have been reported in Figure~\ref{fig:data} alongside predictions of the Kox-Shen semi-empirical model (see section~\ref{sec:formulae}) to guide the reader's eyes. More details about the data sets are given in section~\ref{sec:results}.
	
\begin{figure}[htb]
		\centering
		{\includegraphics[width=.8\textwidth]{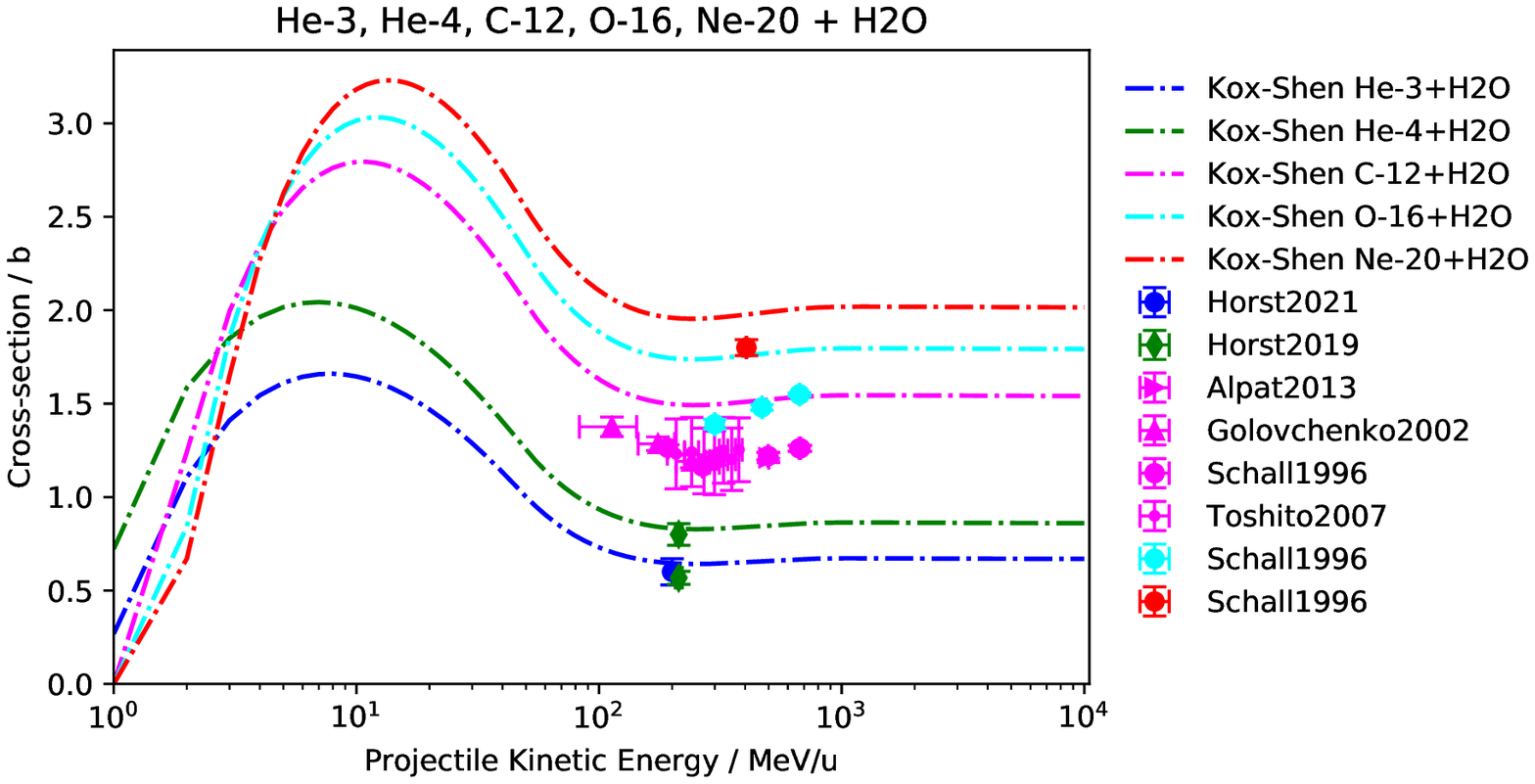}}
		\caption{Data collections for $^3$He, $^4$He, $^{12}$C, $^{16}$O and $^{20}$Ne projectiles on water with the predictions of the Kox-Shen semi-empirical model (see section~\ref{sec:formulae}) to guide the reader's eyes. Different colors represent different projectiles. Both reaction and charge-changing cross-sections are plotted. For $^{4}$He + H$_\textup{2}$O both charge and mass-changing cross-section data were measured by Horst \textit{et al.} \cite{Horst2019,Horst2021}. Mass-changing cross-sections were only measured by Horst \textit{et al.} \cite{Horst2019} for $^3$He and $^4$He. For all other projectiles, the data are charge-changing only. Therefore, the data points lie below the model predictions (see section~\ref{sec:cc_mc}). Data are from references~\cite{Horst2021,Horst2019,Alpat2013,Golovchenko2002,Schall1996}. }
		\label{fig:H2Odata}
	\end{figure}

	Experimentally-measured cross-section data for composite targets (i.e. molecular targets, not made of one element only) are also included in the database. In Figure~\ref{fig:H2Odata}, data for different projectiles impinging on water targets are reported. Projectiles of interest for radiation protection in space and heavy-ion therapy have been chosen. $^{20}$Ne beams were used in the past for heavy-ion therapy \cite{Ne_therapy}.
	
	\subsection{Reaction cross-sections: theoretical background}
	\label{Theory}
	
	During cross-section measurements, a projectile nucleus is shot against a target material. Elastic or inelastic processes can take place between projectile and target nuclei. On the one hand, when elastic interactions happen, no change in the composition of the nuclei takes place and the total kinetic energy of the system is conserved. On the other hand, there are two outcomes of inelastic interactions. The first is projectile and target nuclei staying intact and total kinetic energy not being conserved because of 
	the excitation processes occurring. The second outcome occurs when projectile and target react and either or both of them break apart and produce secondary nuclei. This nuclear reaction process is known as fragmentation and the secondary nuclei are called either projectile or target fragments, depending on which nucleus they originate from. 
	
	The cross-section for a nuclear process to happen is called the reaction cross-section. The cross-section for the production of a specific fragment as consequence of nuclear fragmentation processes, is called the production cross-section. 
	Reaction and production cross-section data can be either integral (total) or differential. In the first case, all the nuclei coming out of the interaction of the projectile with the target material are detected, regardless of their outgoing kinetic energy or angle. Differential cross-sections can be either double-differential if the nuclei are detected at a certain angle and their kinetic energy is resolved, or they can be single-differential. The latter are defined either over the angle, meaning that nuclei are detected at a certain emission angle regardless of their kinetic energy, or over the energy meaning that they are detected for all emission angles, but resolving their kinetic energy.
	This work focuses on total reaction cross-sections only.
	
	Many reaction cross-section data are measured in inverse kinematics (IK), which means that projectile and target have exchanged roles during the measurement. The passage from inverse to direct kinematics is straightforward since the reaction cross-section of e.g. \SI{220}{MeV/u} $^{12}$C (projectile) impinging on $^{27}$Al (target) is the same as the cross-section of \SI{220}{MeV/u} $^{27}$Al (projectile) impinging on $^{12}$C (target). The reason is that having the same kinetic energy per nucleon, both projectiles have the same velocity. Therefore, the center-of-mass energy is the same for the two inverse colliding systems.
	
	\subsection{Charge and mass-changing cross-sections}
	\label{sec:cc_mc}
	
	One of the most fundamental quantities for heavy-ion transport calculations is the total reaction cross-section $\sigma_\textup{R}$ as a function of energy, which provides the probability for a nuclear reaction to occur. However, to measure all kinds of possible reaction channels during an experiment is very difficult and therefore, experimental cross-sections are often only estimates of the total reaction cross-sections. Many experiments measure charge-changing cross-sections $\sigma_\textup{cc}$, which give the probability that the projectile changes its atomic number (``charge''), because a reliable charge identification is rather straightforward with simple particle detection systems. Since most nuclear fragmentation channels lead to loss of at least one proton, the charge-changing cross-section is a good approximation to the reaction cross-section. However, most heavy-ion projectiles can also undergo neutron-removal reactions (for example fragmentation of $\mathrm{^{12}C}$ into $\mathrm{^{11}C}$ or $\mathrm{^{10}C}$), where the charge remains unchanged. The contribution of such reaction channels is not included in the charge-changing cross-section. A quantity that is closer to the total reaction cross-section than the charge-changing cross-section is the mass-changing cross-section $\sigma_\textup{mc}$. It describes the probability for a reaction that changes the nucleus mass number. It also includes such pure neutron-removal channels.
	
\begin{figure}[htb]
		\centering
		\includegraphics[width=0.7\textwidth]{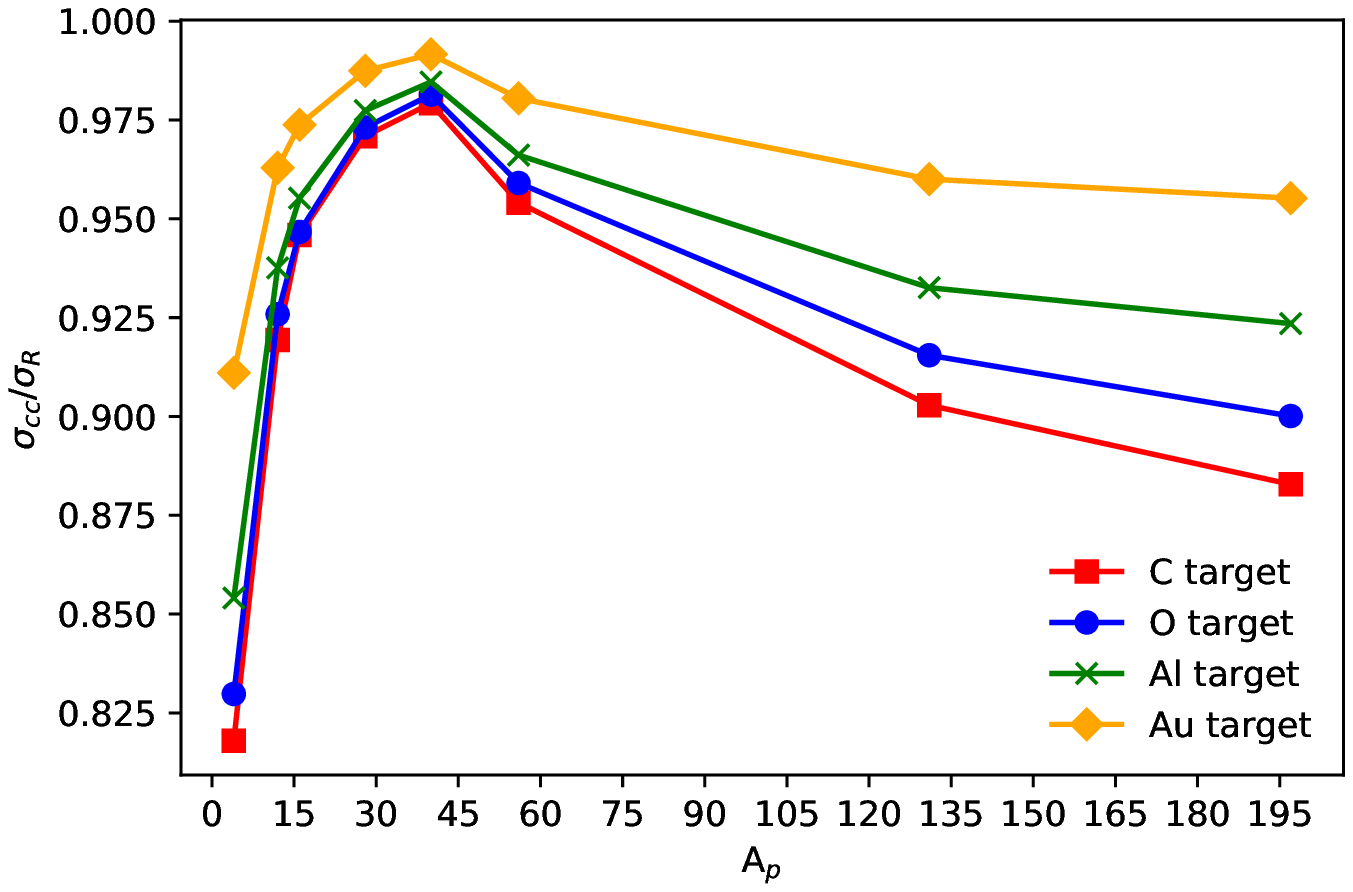}
		\caption{Ratio of charge-changing to total reaction cross-section $\sigma_\textup{cc}/\sigma_\textup{R}$ as a function of the projectile mass number, computed for projectile kinetic energy $1\mathrm{\ GeV/u}$ and for different target materials. The total reaction cross-sections were calculated using the Kox-Shen model \cite{Sihver2014KS} and the neutron-removal cross-sections using the parameterization by Mei \textit{et al.} \cite{Mei2017}.}
		\label{fig:ratio_cc_mc}
	\end{figure}

	The ratio $\sigma_\textup{cc}/\sigma_\textup{R}$ was computed for a variety of colliding systems at $1\mathrm{\ GeV/u}$ (see Figure~\ref{fig:ratio_cc_mc}) with the aim of studying for what colliding systems charge-changing cross-sections can be used to validate total reaction cross-section models and for what systems better estimates are required.  $\sigma_\textup{R}$ was calculated with the Kox-Shen model \cite{Sihver2014KS} and the neutron-removal cross-section ($\sigma_\textup{R}-\sigma_\textup{cc}$) was obtained from the parameterization by Mei \cite{Mei2017} as implemented in the program LISE++ \cite{Kuchera2015}.
	
	A systematic behavior of the $\sigma_\textup{cc}/\sigma_\textup{R}$ ratio can be observed. In the examples shown in Figure~\ref{fig:ratio_cc_mc}, neutron-removal reactions are predicted to have a particularly high contribution for light projectiles (up to $\sim 20 \%$ for $\mathrm{^4He}$). For these projectiles, the number of possible fragmentation channels is very limited and therefore, single neutron-removal has a relatively high probability. Towards heavier projectiles, the $\sigma_\textup{cc}/\sigma_\textup{R}$ ratio increases, which means that the relative probability for neutron-removal reactions decreases.  After reaching a maximum for $\mathrm{^{40}Ca}$ projectiles, the $\sigma_\textup{cc}/\sigma_\textup{R}$ ratio decreases again towards heavier nuclei. This can be explained by their neutron excess. Up to $\mathrm{^{40}Ca}$, stable nuclei have the same number of neutrons and protons while above they consist of more neutrons than protons. The higher relative abundance of neutrons in heavier nuclei makes also neutron-removal reactions more probable again. The different curves show the dependence on the target material. It can be observed that towards heavier targets, the contribution of neutron-removal reactions to the total reaction cross-section decreases. This trend can be explained by the fact that peripheral collisions, which are the main cause for the removal of single nucleons, are more probable for light target nuclei.
	
	Nevertheless, these calculations only predict an estimate of the real cross-section ratio. The few experimental data that exist for intermediate systems are systematically underestimated \cite{Bochkarev1998,Webber1990,Westfall1979}.
	Since charge-changing cross-sections can have large deviations from reaction cross-sections, only reaction and mass-changing cross-section data are compared to model predictions in section~\ref{sec:results}.
	The only exception are $^3$He data. Charge and mass-changing cross-sections for the case of $^3$He are identical because there are no pure neutron-removal channels. 
	
	\subsection{Database entries}
	\label{sec:DB}
	
	The database is presented as a table with the following columns:
	\begin{itemize}
		\item Projectile atomic number 
		\item	Projectile mass number
		\item	Target atomic number. For compound materials (e.g. water or methane) the effective atomic number has been computed as \cite{Mayneord1937}:
		\begin{equation}
			Z_\textup{eff} = \left( { \sum_k f_k (Z_k)^{2.94} } \right)  ^ {\frac{1}{2.94}},
			\label{Eq:Z_eff}
		\end{equation}
		where $f_k$ is the fraction of the total number of electrons associated with each element and $Z_k$ is the atomic numbers of the element. 
		\item	Target mass number. The sum of the single mass numbers of each element is reported in compound materials. If the target is described just as an element without specifications about the isotope, it is reported with the standard atomic weight rounded to the closest integer (e.g. 64 for Cu).
		\item	Target chemical formula. This is useful for the case of compound targets.
		\item	Target areal density (g/cm$^2$). This is an important parameter to evaluate the quality of the data: thinner targets give better data as the projectile kinetic energy during the reaction is better defined and the probability of multiple reactions is lower. However, thinner targets increase the beamtime necessary to collect data with appropriate statistical uncertainty \cite{Norbury2020}. If more than one target was used, more than one value is reported.
		\item	Projectile kinetic energy (MeV/u)
		\item	Projectile kinetic energy lower uncertainty (MeV/u)
		\item	Projectile kinetic energy upper uncertainty (MeV/u)
		\item	Evaluation point of the projectile energy. The projectile energy reported in the publication could be either the primary energy from the accelerator (e.g. identified as ‘‘out of the beamline'' or ‘‘before the target'') or the energy at the center of the target, especially for thick targets (e.g. identified as ‘‘in the center of the target'').
		\item	Cross-section type. The three main cross-section types are charge-changing (‘‘cc''), mass-changing (‘‘mc'') and reaction. A charge-changing cross-section is defined as the probability that the projectile nucleus loses at least one proton (this does not include neutron-removal reactions).  A mass-changing cross-section is the probability that the projectile loses at least one nucleon (this includes neutron-removal reaction channels). For heavy-ion reactions, the reaction cross-section is normally almost identical to the mass-changing cross-section. It is to be noticed that it is more difficult to measure reaction cross-sections for heavy projectile nuclei. Therefore, the number of reaction cross-section data (green contributions in Figure~\ref{fig:counts}) become lower with increasing projectile atomic number and mostly charge-changing cross-sections are available.
		Since the definition of what the authors mean with a specific cross-section type is publication dependent, it is recommended to look into the specific work for a deeper understanding.
		\item	Cross-section (mb). This is the cross-section value reported in millibarn.
		\item	Cross-section lower uncertainty (mb)
		\item	Cross-section upper uncertainty (mb)
		\item	Uncertainty type. This can be either purely statistical or both statistical and systematic. It is important to notice that, especially in older publications, the reported uncertainties often do not include systematic components due to e.g. instrument calibration, but only include the statistical component. 
		When using data whose uncertainty evaluation only includes the statistical component, it should be considered that the error bars are actually larger than they appear.
		\item	First author of the publication
		\item	Year of the publication
		\item	DOI: unique Digital Object Identifier of the peer-reviewed publication the data come from
		\item	Experiment facility
		\item	Beamtime. If the month and the year of the experiment are reported in the publication, they are also reported in the database.
		\item	Measurement method. The detectors used to experimentally obtain the data have sometimes been added as well. It is recommend to refer to the publications for further details. 
		\item	Comments: other comments.
	\end{itemize}
	For a few cases, electromagnetic dissociation (EMD) cross-section data are presented in the publications. They have also been reported in the database (in mb) alongside the lower and upper uncertainties (mb) associated to them. However, EMD processes are not included in the presented total reaction cross-section parameterizations. They should be modeled separately because EMD cross-sections follow scaling laws other than nuclear fragmentation and also the reaction products can differ.
	
	\section{Total nuclear reaction cross-section parameterizations}
	\label{sec:formulae}
	
	Several semi-empirical parameterizations were developed over the last forty years to describe the trend of total reaction cross-sections as a function of the energy of the projectile-target system. Within this work, the Kox \cite{Kox1987}, Shen \cite{Shen1989}, Kox-Shen \cite{Sihver2014KS}, Tripathi \cite{Tripathi1996,Tripathi1999} with Horst optimizations for $^4$He projectiles \cite{Horst2019} and Hybrid-Kurotama \cite{Sihver2014HK} semi-empirical parameterizations were re-implemented in the Python programming language. Details about the implementation of the formulae can be found in the corresponding sections. 
	
	Kox, Shen and Tripathi parameterizations are implemented in Geant4 and can be used, even though none of them is used by default within any of the Geant4 physics lists. Hybrid-Kurotama is the default parameterization for PHITS (Kox-Shen and Tripathi are options). An empirically-modified version of Tripathi is used in FLUKA and the optimizations introduced by Horst \textit{et al.} for $^4$He were implemented in a similar form in the last version of FLUKA 
	\cite{Arico2019}.
	The deterministic codes GSI in-house heavy-ion treatment planning system TRiP98 \cite{Kraemer2000} and its extension for space SpaceTRiP \cite{SpaceTRiP}, also use Tripathi corrected with Horst optimizations as default for total cross-sections. 
	
	Most of the nuclear reaction cross-section semi-empirical parameterizations are based on the geometrical Bradt-Peters approach \cite{Bradt-Peters1950}. It estimates the cross-section as:
	\begin{equation}
		\sigma_R = \pi {r_0}^2 ( {A_p}^{1/3} + {A_T}^{1/3} + \delta  )^2.
		\label{Eq:sigma_BP}
	\end{equation}
	$A_p$ and $A_T$ are, respectively, the projectile and target mass numbers, $r_0$ is the nucleon radius and $\delta$ is the overlap transparency parameter.

	\subsection{Kox parameterization}
	\label{sec:K}
	
	The phenomenological Kox parameterization was proposed in 1987~\cite{Kox1987}:
	\begin{equation}
		\sigma_R = \pi {R_\textup{int}}^2  \, \biggl( 1 - \frac{B}{E_\textup{cm}} \biggr).
		\label{Eq:sigma_Kox}
	\end{equation}
	$B$ is the Coulomb interaction barrier and it is energy independent. $R_\textup{int}$ is the interaction radius and it has three components. The first is similar to the Bradt-Peters interaction radius ($R_\textup{vol}$). The second and the third are the so-called ``nuclear surface'' contribution ($R_\textup{sur}$) and the neutron excess ($D$), which is explained to be necessary only for projectile kinetic energies below \SI{200}{MeV/u} in Ref.~\cite{Kox1987}. The $R_\textup{sur}$ term accounts for mass asymmetry (energy independent) and transparency. The transparency parameter is called $c$. Some tabulated values of $c$ are proposed within Table III of Ref.~\cite{Kox1987} between \SI{30}{} and \SI{2100}{MeV/u}. 
	$E_\textup{cm}$ is the center-of-mass energy of the system in \si{MeV} calculated as:
	\begin{equation}
		E_\textup{cm} = E \frac{A_p A_T}{A_p + A_T},
		\label{Eq:E_cm}
	\end{equation}
	with $E$ being the kinetic energy of the projectile in \si{MeV/u}. Equation~\ref{Eq:E_cm} is not exact but it is a good approximation for the following reason. At low energies, the values of $E_\textup{cm}$ extracted from Equation~\ref{Eq:E_cm} are identical to the values that are obtained with the proper calculation. When it comes to high energies, the properly-computed $E_\textup{cm}$ and the approximated value obtained with Equation~\ref{Eq:E_cm} become more and more different, but the ratio $B / E_\textup{cm}$ becomes small and $1 - (B / E_\textup{cm}) \simeq 1$.
	
	The model has been implemented within the present work as in Geant4. 
	This means that the following exceptions have been applied to the description of Ref.~\cite{Kox1987}:
	\begin{enumerate}
		
		\item The neutron excess parameter $D$ has been used for all kinetic energies, not only below \SI{200}{MeV/u}.
		
		\item The following functions have been used for the transparency parameter $c$ instead of tabulated values:
		\begin{equation}
			c= \left( - \frac{10}{1.5^5} + 2 \right) \left(\frac{x}{1.5}\right)^3  \qquad \textup{for} \quad x<1.5
			\label{Eq:c_Kox_G4_1}
		\end{equation}
		\begin{equation}
			c= - \frac{10}{x^5} + 2  \qquad \qquad \qquad \quad \, \textup{for} \quad x\ge1.5
			\label{Eq:c_Kox_G4_2}
		\end{equation}
		where $x=\log_{10} (E)$. 
		
		\item Equation 14 of Ref.~\cite{Kox1987} describes the Coulomb interaction barrier parameter as:
		\begin{equation}
			B = \frac{Z_T  Z_p e^2}{r_C ({A_T}^{1/3} + {A_p}^{1/3})}.
			\label{Eq:B_Kox}
		\end{equation}
		Nevertheless, it not necessary to multiply the projectile and target atomic numbers $Z_p$ and $Z_T$ by the elementary charge $e$ and the expression for $B$ to be used is:
		\begin{equation}
			B = \frac{Z_T  Z_p}{r_C ({A_T}^{1/3} + {A_p}^{1/3})}.
			\label{Eq:B_Kox_corrected}
		\end{equation}
		It is to be noted that $r_C$ ($=\SI{1.3e-15}{m}$) has to be inserted in fm in Equation~\ref{Eq:B_Kox_corrected}, while $r_0$ ($= \SI{1.1e-15}{m}$) in m in all the components of $R$ ($R_\textup{int}$, $R_\textup{vol}$, $R_\textup{sur}$). In Ref.~\cite{Kox1987}, it is not specified that the $D$ parameter in $R$ should also be multiplied by $r_0$. In this way the result is obtained in $\si{m^2}$.
		
		\item A low-energy check has been added. It consists in automatically setting the cross-section to zero for energy values $E_\textup{cm} \le B$.
	\end{enumerate}

	\subsection{Shen parameterization}
	\label{sec:KS}
	
	With the parameters presented in~\cite{Kox1987}, the Kox formula fails in reproducing the data for energies lower than \SI{10}{MeV/u} and no values for $c$ are given below \SI{30}{MeV/u}. For these reasons, in 1989 a new unified parameterization based on the Kox's formula was proposed by Shen \textit{et al.}~\cite{Shen1989}. It can be used for both low  and intermediate energy ranges. The Shen formula for the cross-section is the same as Equation~\ref{Eq:sigma_Kox}, but the $B$ and $R_\textup{int}$ parameters are different. In equation 5 of Ref.~\cite{Shen1989}, the use of the following $B$ parameter is proposed:
	\begin{equation}
		B = 1.44 \, \frac{Z_T  Z_p}{R_T + R_p + 3.2} - b \, \frac{R_T  R_p}{R_T + R_p},
		\label{Eq:B_Shen}
	\end{equation}
	where:
	\begin{equation}
		R_i = 1.12 \, {A_i}^{1/3} - 0.94 \, {A_i}^{-1/3} \quad \quad \quad i=(T,p)
		\label{Eq:R_i_Shen}
	\end{equation}
	and $b=1$. As pointed out in Equation~\ref{Eq:R_i_Shen}, $i$ can stand either for target or projectile.
	In the Shen model, the $R_\textup{int}$ parameter is the same as Kox but with two modifications. Firstly, an energy-dependent term was added. 
	Secondly, the multiplication factor called $\alpha$ of the neutron excess term changes from being equal to 5, as recommended for the Kox model, to being equal to 1. 
	Nevertheless, it is pointed out that by using $\alpha=1$, the cross-sections for heavy targets are underestimated. 
	In Figure~2 of Ref.~\cite{Shen1989}, the $c$ parameter recommended to be used for $R_\textup{int}$ is reported. For energies above \SI{30}{MeV/u}, it is the same as in~\cite{Kox1987}. 
	However, no numerical or functional formula for $c$ is provided in Ref.~\cite{Shen1989}.
	In Ref.~\cite{Sihver2014KS}, it is explained that $c$ values from the curve in Figure~2 of Ref.~\cite{Shen1989} should not be extracted because they are not fully in agreement with the $c$ parameter curve reported in Figure~12 of Ref.~\cite{Kox1987} and because the energy scale of the figure is inconsistent.
	For this reason, $c$ from equations~\ref{Eq:c_Kox_G4_1} and~\ref{Eq:c_Kox_G4_2} has been used in the current work, as implemented in Geant4.
	Also in this case, the same low-energy check used for the Kox formula has been implemented for the Shen parameterization.
	
	The Kox and Shen parameterizations are planned to be removed from the next Geant4 version.

	\subsection{Kox-Shen parameterization}
	\label{sec:KSS}
	
	Recently, a modified parameterization for the transparency parameter $c$ of the Shen model has been proposed~\cite{Sihver2014KS}:
	\begin{equation}
		c= \left( - \frac{10}{1.5^5} + 2 \right) \left(\frac{x}{1.38}\right)^3 + 0.0006 \, E  \qquad \qquad \textup{for} \quad E \le \SI{45}{MeV/u} 
		\label{Eq:c_KSS_1}
	\end{equation}
	\begin{equation}
		c= 1.91 - 16 \, \textup{e} ^{-0.7274 \, E^{0.3493} }  \cos(0.0849 \, E^{0.5904}) \quad \textup{for} \quad E > \SI{45}{MeV/u}
		\label{Eq:c_KSS_2}
	\end{equation}
	$E$ is the kinetic energy of the projectile in \si{MeV/u} and $x=\log_{10} (E)$. The expression valid for $E > \SI{45}{MeV/u}$ was already proposed in 1988 by Townsend and Wilson~\cite{Townsend1988}. The expression for $E \le \SI{45}{MeV/u}$ is very similar to the one developed for Geant4, but with some modifications that provide a smooth overlap with the Townsend and Wilson's part.
	
	In addition, the suggestion of always using the lighter particle as projectile is provided in Ref.~\cite{Sihver2014KS}.
	
	The value $\alpha=5$ is used for the neutron-excess parameter multiplication factor in Ref.~\cite{Sihver2014KS}. However, as explained in Ref.~\cite{Shen1989}, the $\alpha$ value that fits the experimental data best is 1. For this reason, $\alpha=1$ has been used in the implementation of this work.

	\subsection{Tripathi parameterization}
	\label{sec:Tri}
	
	The Tripathi semi-empirical formula was first presented in 1996 \cite{Tripathi1996}. Two publications followed in 1997 \cite{Tripathi1997} and 1999 \cite{Tripathi1999}. They respectively dealt with neutron projectiles and light systems, where ``light systems'' means that at least either projectile or target has mass number $A \le 4$. The parameterization for neutron projectiles \cite{Tripathi1997} has not been included in the present work as neutron data are not part of the data collection. The other two parameterizations will be referred to as ``Tripathi96'' and ``Tripathi99'' in the following.
	
	\subsubsection{Tripathi96 parameterization}
	\label{sec:Tri96}
	
	The following form for the reaction cross-section was presented in Ref.~\cite{Tripathi1996}:
	\begin{equation}
		\sigma_R = \pi {r_0}^2 ( {A_p}^{1/3} + {A_T}^{1/3} + \delta_E  )^2 \, \biggl( 1 - \frac{B}{E_\textup{cm}} \biggr) f,
		\label{Eq:sigma_Tri96}
	\end{equation}
	where $r_0 = \SI{1.1}{fm}$, $B$ is the energy-dependent Coulomb barrier and $f$ is a multiplication factor equal to $1$ in all cases but for $^1$H + $^4$He and $^1$H + $^{12}$C, where it is supposed to be set to 27 and 3.5, respectively. 
	Also in this case, Equation~\ref{Eq:E_cm} is used for the computation of $E_\textup{cm}$.
	In Geant4, the proper physical calculation for $E_\textup{cm}$ is implemented, while Equation~\ref{Eq:E_cm} is used in PHITS.
	
	The parameter $\delta_E$ is defined as:
	\begin{equation}
		\delta_E = 1.85 \, S + 0.16 \frac{S} {{E_\textup{cm}}^{1/3} } - C_E + \alpha \frac{ (A_T - 2 Z_T) Z_p } {A_T A_p}.
		\label{Eq:d_E}
	\end{equation}
	The last term is commonly called the neutron excess parameter and the multiplication factor is $\alpha=0.91$. $S$ is the mass asymmetry term and $C_E$ is the parameter through which $\delta_E$ accounts for the transparency and Pauli-blocking effects. $C_E$ itself is energy dependent and can be computed as:
	\begin{equation}
		C_E = D ( 1 + \exp(-E/T_1)) - 0.292 \exp(-E/792) \cos(0.229 E^{0.453}),
		\label{Eq:C_E}
	\end{equation}
	where $T_1=40$, $E$ is the projectile kinetic energy in \si{MeV/u} and $D$ is proportional to the density of the colliding system, scaled with respect to the density of the system $^{12}$C+$^{12}$C:
	\begin{equation}
		D = 1.75 \, \frac{\rho_{A_p} + \rho_{A_T}}{\rho_{A_C} + \rho_{A_C}}.
		\label{Eq:D}
	\end{equation}
	More details can be found in Ref.~\cite{Tripathi1996}. Moreover, in Ref.~\cite{Tripathi1996} it is recommended to use: 
	\begin{itemize}
		\item the single value $D=2.05$ for the proton - nucleus case
		\item the reduced value $D/3$ for lithium nuclei
		\item the specific density-independent formula for the case of $^4$He projectiles, due to the small density compression:
		\begin{equation}
			D = 2.77 - \SI{8.0e-3}{} A_T + \SI{1.8e-5}{} {A_T}^2 - \frac{0.8}{1 + e^{\frac{250-E}{G}}},
			\label{Eq:D_He_Tri96}
		\end{equation}
		where $G=75$. It is believed that there is a typo in the original publication \cite{Tripathi1996} concerning the parentheses for $D$, since Equation~\ref{Eq:D_He_Tri96} is consistent with the formula given for $D$ in Ref.~\cite{Tripathi1999}. 
	\end{itemize}  
	Tripathi96 has been implemented in the present work as described in Ref.~\cite{Tripathi1996}, with a few modifications:
	\begin{enumerate}
		\item A different nuclear radius $r_i$ has been used. The nuclear radius $r_i$ is inside the Coulomb barrier parameter $B$. In Ref.~\cite{Tripathi1996} it is suggested to use:
		\begin{equation}
			r_i =1.29 \, r_{\textup{rms},i}
			\label{Eq:r_i}
		\end{equation}
		and to use data from Ref.~\cite{DeVries1987} for  $r_{\textup{rms},i}$. This will be called ``Wilson $r_i$''. 
		In particular, the $r_{\textup{rms},i}$ used for the present work is the arithmetic average of the data given in Ref.~\cite{DeVries1987} for $Z \le 26$ and $r_{\textup{rms},i} = 0.84 \, \,  {A_i}^{1/3} + 0.55$ for $Z > 26$. The formula for $Z > 26$ is reported in Appendix A of Ref.~\cite{Tripathi1999WR}. 
		Nevertheless, in Geant4 \cite{Geant4_UserManual}, the following formula is used to compute $r_i$ of projectile and target nuclei:
		\begin{equation}
			r_i = \frac{1.29 \times 0.6 \times \SI{1.36e-15}{} \, {A_i}^{1/3} } { r_0 }.
			\label{Eq:r_i_G4}
		\end{equation}
		In Figure~\ref{fig:r_nuclear}, the radius computed as in Equation~\ref{Eq:r_i_G4} is referred to as ``G4 $r_i$''.
		It was observed that the Tripathi model is sensitive even to small changes of this parameter at low energies. 
		It has been decided to use for $r_i$ the same formula as implemented in Geant4 (Equation~\ref{Eq:r_i_G4}) after comparing all the low energy ($\le$ \SI{10}{MeV/u}) cross-sections from the data collection with the results obtained either with Tripathi96 implemented with $r_i$ from Equation~\ref{Eq:r_i} or Equation~\ref{Eq:r_i_G4} (see Figure~\ref{fig:r_nuclear}). The G4 $r_i$ does not fit the experimental data best for all systems. However, the decision of using it in the implementation comes from the following considerations. For the systems: $^4$He + $^{12}$C (``Labie1973'' dataset~\cite{Labie1973}), $^4$He + $^{27}$Al (one data point),  $^4$He + $^{56}$Fe (two data points) and $^4$He + $^{237}$Np (``Powers1966'' dataset~\cite{Powers1966}), G4 $r_i$ fits the data best. For $^4$He + $^{181}$Ta and $^4$He + $^{197}$Au, the Wilson $r_i$ applied to Tripathi96 fits the data best. For $^{12}$C + $^{12}$C, both Wilson and G4 $r_i$ are compatible with the single data point. For the cases of $^4$He + $^{181}$Ta and $^4$He + $^{197}$Au, the measurements are only single data points, while for $^4$He + $^{237}$Np there is a series of measurement points that systematically follow the cross-section increase in the Coulomb barrier energy region (see Figure~\ref{fig:r_nuclear}). In PHITS, the Wilson radius is used.
		
		\begin{figure}[htb]
			\centering
			\subfloat[][]
			{\includegraphics[width=.55\textwidth]{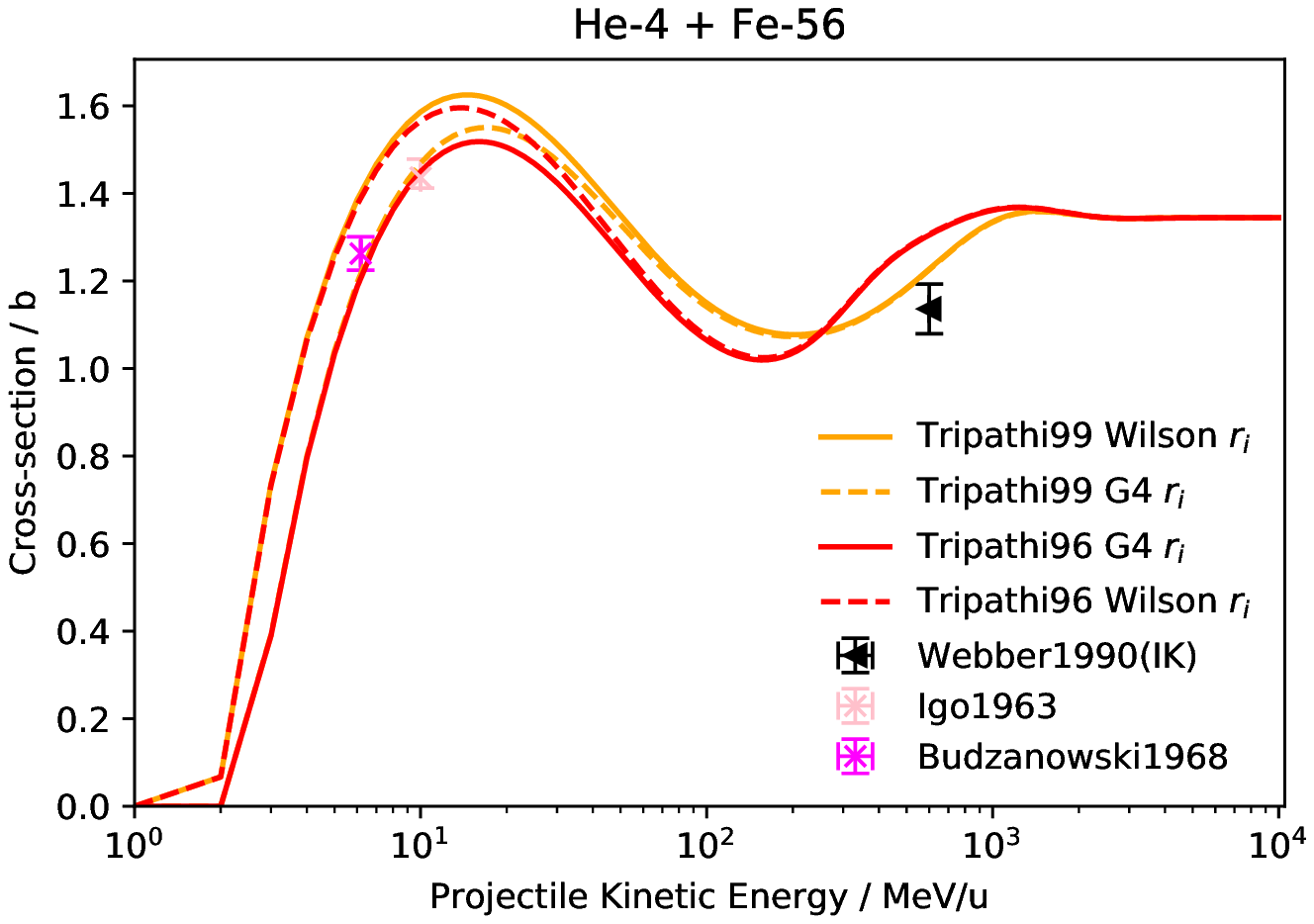}} 
			\subfloat[][]
			{\includegraphics[width=.55\textwidth]{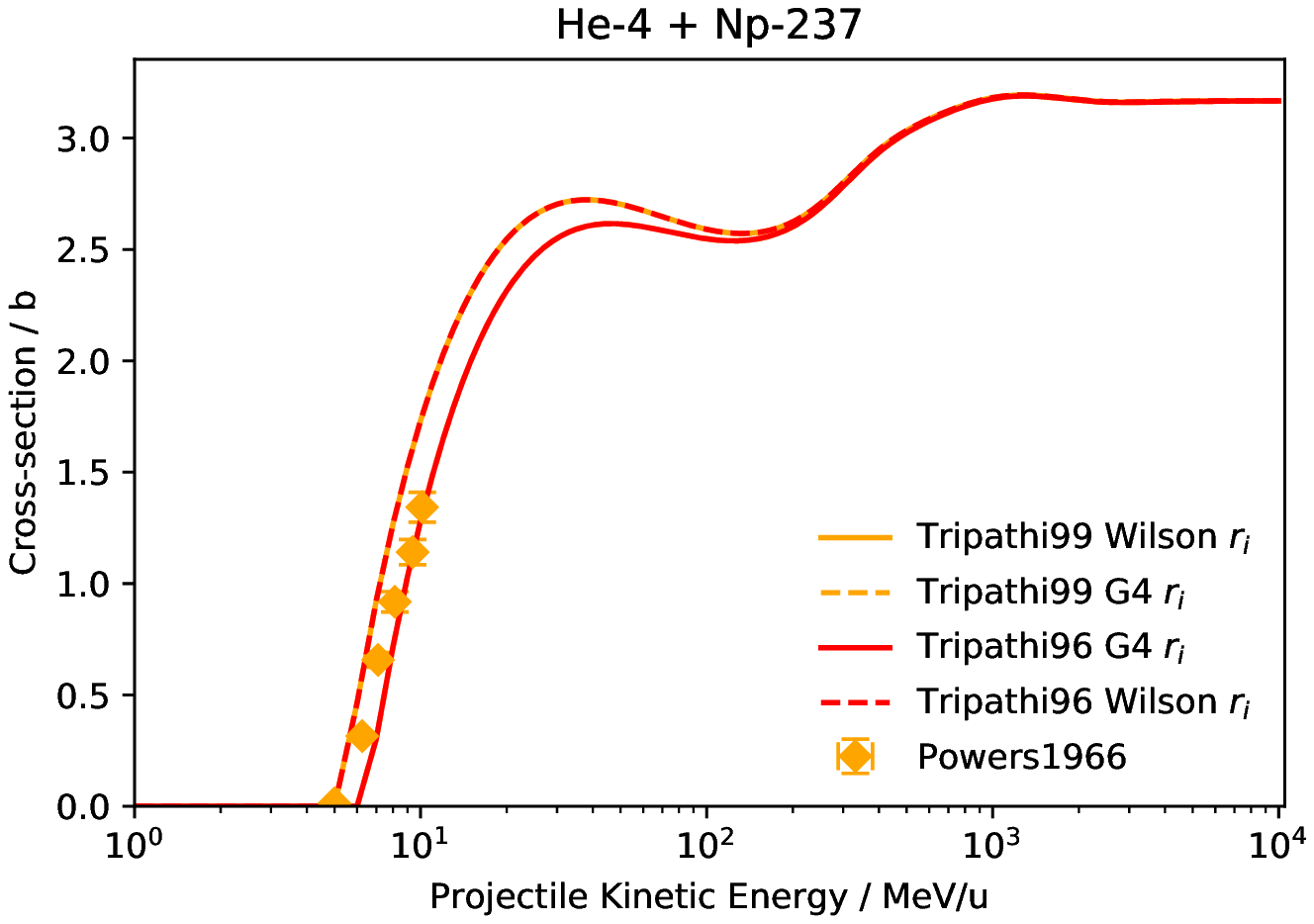}}
			\caption{Dependence of Tripathi96 and Tripathi99 semi-empirical cross-section formula on the nuclear radius for $^4$He + $^{56}$Fe and $^4$He + $^{237}$Np. Solid lines have been chosen for the version of the models that has been implemented within the present work. Data in panel (a) are from references~\cite{Webber1990,Igo1963,Budzanowski1967}, in panel (b) from reference~\cite{Powers1966}. }
			\label{fig:r_nuclear}
		\end{figure}
		
		\item Horst \textit{et al.} optimizations \cite{Horst2019} to Equation~\ref{Eq:D_He_Tri96} are used to calculate $D$ for $^4$He + targets from C to Si.
		Recently some charge- and mass-changing cross-section measurements with $^4$He were performed at therapeutic energies (\SI{70}{} - \SI{220}{MeV/u}) on $^{12}$C, $^{16}$O and $^{28}$Si targets \cite{Horst2017,Horst2019}. Based on these data and the data by Ingemarsson \textit{et al.} \cite{Ingemarsson2000}, an optimization of Tripathi96 for the case of $^{4}$He projectiles on targets with masses between C and Si has been proposed being:
		\begin{equation}
			D = 2.2 - \SI{8.0e-3}{} A_T + \SI{1.8e-5}{} {A_T}^2 - \frac{0.3}{1 + e^{\frac{120-E}{G}}}.
			\label{Eq:D_Horst}
		\end{equation}
		where $G=50$. These changes led to considerable improvements of $^{4}$He dose calculations \cite{Horst_PhDThesis,Arico2019}.
		
		\item Low-energy check: once cross-sections are computed for all energies, any negative values are set to null. This procedure is implemented in Geant4 as well.
		
	\end{enumerate} 
	
	In Tripathi96 calculations performed for the present work, the lightest ion is always considered to be the projectile. This changes the choice of $D$ parameter to be used and the neutron excess parameter (last term of Equation~\ref{Eq:d_E}). The same is done in Geant4 and PHITS.

	\subsubsection{Tripathi99 parameterization}
	\label{sec:Tri99}
	The Tripathi99 formula was implemented as presented in Ref.~\cite{Tripathi1999}:
	\begin{equation}
		\sigma_R = \pi {r_0}^2 ( {A_p}^{1/3} + {A_T}^{1/3} + \delta_E  )^2 \, \biggl( 1 - R_c \frac{B}{E_\textup{cm}} \biggr) X_m.
		\label{Eq:sigma_Tri99}
	\end{equation}
	The additional terms with respect to Equation~\ref{Eq:sigma_Tri96} are the system-dependent Coulomb multiplier $R_c$, which allows one to keep the same formalism for light, medium and heavy nuclei, and the optical model multiplier $X_m$:
	\begin{equation}
		X_m = 1 - X_1 \textup{e}^{-\frac{E}{X_1 \, S_L}},
		\label{Eq:X_m}
	\end{equation}
	where $X_1=5.2$ for the n+$^4$He system and
	\begin{equation}
		X_1 = 2.83 - \SI{3.1e-2}{} A_T + \SI{1.7e-4}{} {A_T}^2
		\label{Eq:X_1}
	\end{equation}
	in all other cases, and
	\begin{equation}
		S_L = 1.2 + 1.6 (1 - \textup{e}^{-\frac{E}{15}}).
		\label{Eq:S_L}
	\end{equation}
	Differently from Tripathi96, $T_1$ and $G$ are system dependent in Tripathi99. The nuclear radius used in this case is the Wilson $r_i$ (Equation~\ref{Eq:r_i}), as recommended in Ref.~\cite{Tripathi1999}. Also in Geant4 and PHITS the Wilson $r_i$ is used for Tripathi99. To be noticed is that the Geant4 radius would fit the experimental data better for all systems for which low-energy data were measured (see section~\ref{sec:Tri96} and Figure~\ref{fig:r_nuclear}).
	From Ref.~\cite{Tripathi1999}, the lightest particle is to be used as the projectile in the formulation. This is how the model is implemented in Geant4, PHITS and the present work.
	The model has been implemented as in Ref.~\cite{Tripathi1999}, with a few modifications:
	\begin{enumerate}
		
		
		\item The center-of-mass kinetic energy of the system $E_\textup{cm}$ is in \si{MeV}. We believe that the unit of measurement given for it in Ref.~\cite{Tripathi1999} (\si{A MeV}) is a typographical error.
		
		\item In the Tripathi subroutine of PHITS, $X_m=1$ is used for every projectile ion but neutrons. 
		Using $X_m=1$ instead of $X_m$ from Equation~\ref{Eq:X_m} gives in fact, a better agreement with the original curves presented in Ref.~\cite{Tripathi1999} (the difference is appreciable for all figures from 3 to 20 of Ref.~\cite{Tripathi1999} but 4, 5 and 18).
		In addition, the curves were compared with the measured cross-sections from the database and in the majority of the cases the use of $X_m=1$ gives better agreement with the data. 
		Figure~\ref{fig:X_m=1} shows how the use of $X_m=1$ instead of $X_m$ from Equation~\ref{Eq:X_m}, improves the fit of the model to the data and also to the curve reported in Ref.~\cite{Tripathi1999}.
		
		\begin{figure}[htb]
			\centering
			\subfloat[][]
			{\includegraphics[width=.55\textwidth]{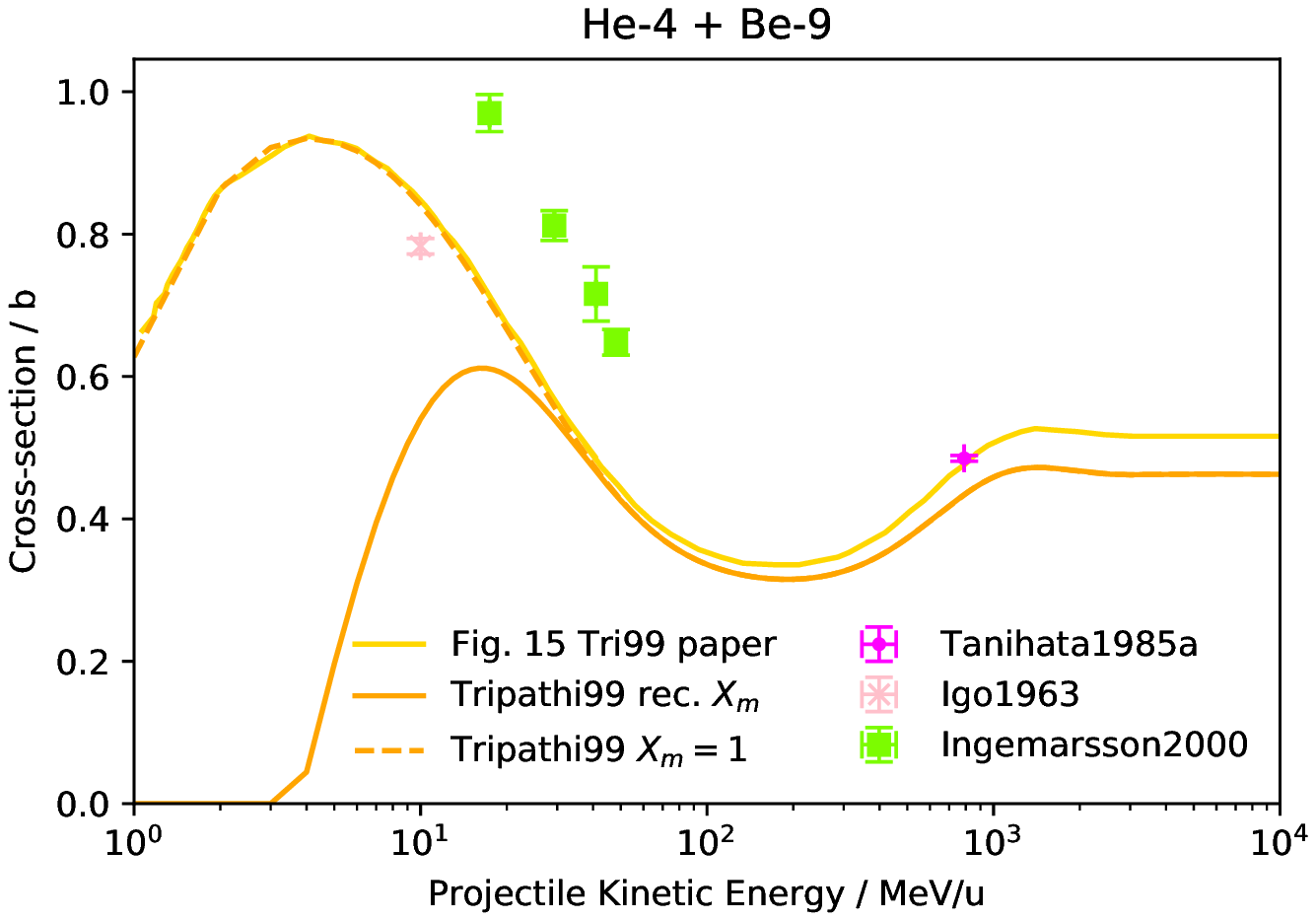}} 
			\subfloat[][]
			{\includegraphics[width=.55\textwidth]{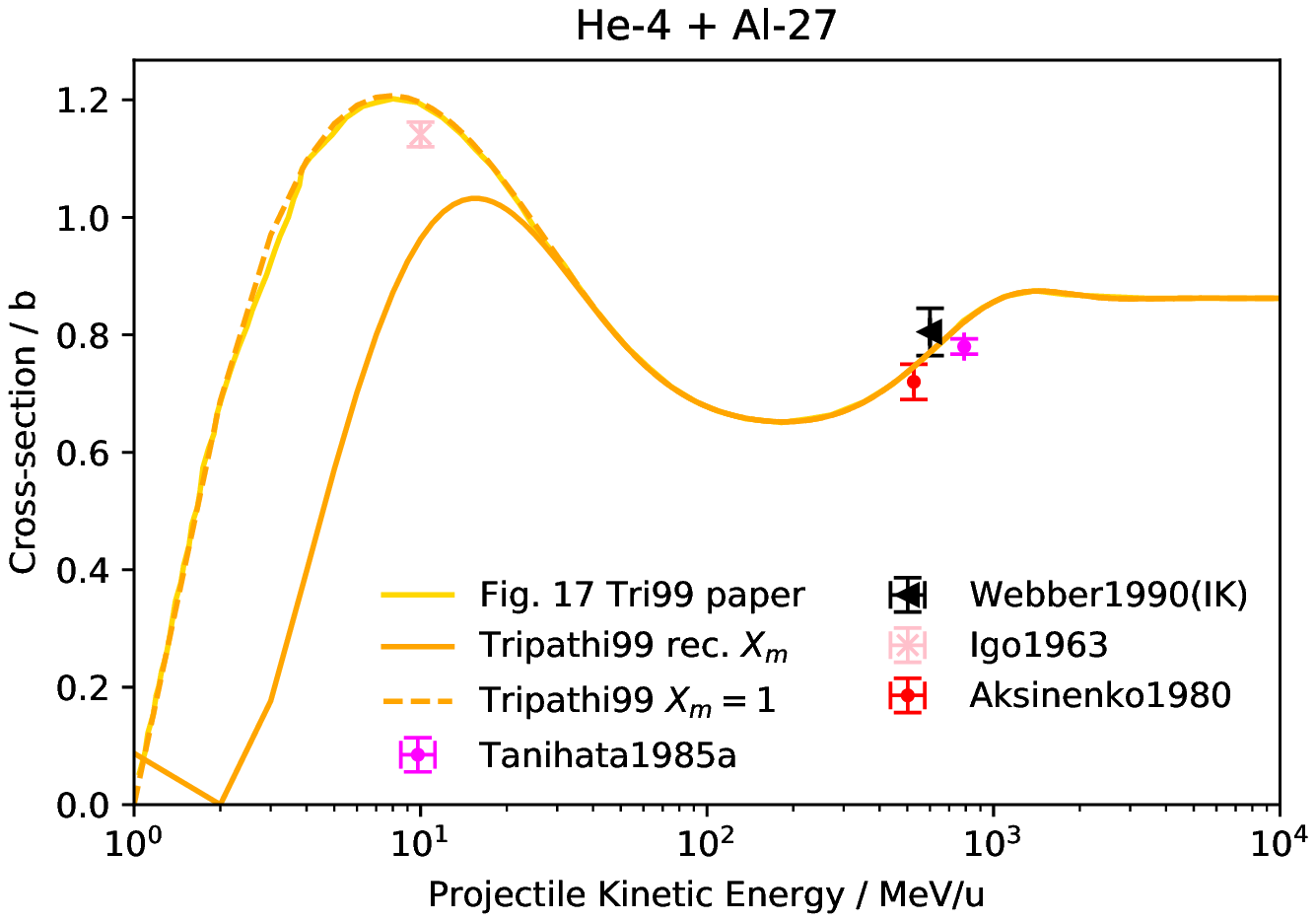}}
			\caption{Comparison between the results of Tripathi99 for the systems $^{4}$He+$^{9}$Be and $^{4}$He+$^{27}$Al, obtained using $X_m=1$ or $X_m$ from Equation~\ref{Eq:X_m}, which is recommended in Ref.~\cite{Tripathi1999} (``rec. $X_m$''), and the curves presented in Ref.~\cite{Tripathi1999}. The experimental data from the database have been plotted as well. For the case of $X_m$ from Equation~\ref{Eq:X_m}, the low-energy check implemented in Geant4 has been used: below \SI{6}{MeV/u}, if the first derivative of the cross-section as a function of the energy is negative (i.e. values becoming smaller with increasing energy) the cross-section values are set to zero. 
			IK stands for Inverse Kinematic data. Data in panel (a) are from references~\cite{Tanihata1985_1,Igo1963,Ingemarsson2000}, in panel (b) from references~\cite{Tanihata1985_1,Webber1990,Igo1963,Aksinenko1980}.
			}
			\label{fig:X_m=1}
		\end{figure}
		
		\item In the Tripathi subroutine of PHITS, optimized $T_1$ and $G$ parameters for a broader set of systems are specified. In particular, specific values are used for the cases of $^{1}$H + $^{3}$He, $^{4}$He, $^{6}$Li, $^{7}$Li, $^{1}$H + targets with $A_T > 7$ and $^{2}$H + $^{4}$He. The values are presented in table~\ref{tab:Tri99PHITS}. For $^{1}$H + other targets, the values used are the same as presented in the original work for $^{1}$H + any target. These additional optimizations have been implemented in the calculation of Tripathi99 for the present work.
		\begin{table}
			\caption{System-dependent values for $T_1$ and $D$ used in the Tripathi subroutine of PHITS and in the present work.}
			\label{tab:Tri99PHITS}
			\centering
			\begin{indented}
				\item[
				\begin{tabular}{lcc}
					\toprule
					Projectile + Target & $T_1$ & $D$ \\
					\midrule
					$^{1}$H + $^{3}$He & 58 & 1.70 \\
					$^{1}$H + $^{4}$He & 40 & 2.05 \\
					$^{1}$H + $^{6}$Li & 40 & 2.05 \\
					$^{1}$H + $^{7}$Li & 37 & 2.15 \\
					$^{1}$H + ($A_T > 7$) & 40 & 2.05 \\
					$^{2}$H + $^{4}$He & 23 & $1.65 + 0.22 / \left(1+\exp{\frac{500-E}{200}}\right)$ \\
					\bottomrule
				\end{tabular}
				]
			\end{indented}
		\end{table}	
	\end{enumerate}
	In Geant4, in addition to the check for negative cross-section values, which is recommended in Ref.~\cite{Tripathi1999}, an extra check was added for Tripathi99. At low energies (below \SI{6}{MeV/u}), if the first derivative of the cross-section as a function of the energy is negative (i.e. cross-section values becoming smaller with increasing energy) the cross-section values are set to null. This check is nevertheless not necessary if $X_m=1$ is used. For this reason, it has not been implemented within the current work.

	\subsection{Hybrid-Kurotama parameterization}
	\label{sec:HK}
	
	A semi-empirical parameterization model called ``Hybrid-Kurotama'' was proposed in 2014 within Ref.\cite{Sihver2014HK}. It is based on the Black Sphere (``Kurotama'' in Japanese) cross-section formula, extended to low energies by smoothly connecting it to the Tripathi parameterization:
	\begin{equation}
		\sigma_R (E) = f_{cut1} \pi ( a_p (E) + a_T (E) )^2 + f_{cut2} (E) \sigma_\textup{Trip} (E) \frac{ \pi ( a_p (E_{cut}) + a_T (E_{cut}) )^2 }{\sigma_\textup{Trip} (E_{cut})}.
		\label{Eq:sigma_HK}
	\end{equation}
	$a_p$ is the black-sphere radius of the projectile, $a_T$ is the black-sphere radius of the target, and $E$ is the projectile kinetic energy in \si{MeV/u}. $E_{cut} = \SI{400}{MeV/u}$ in the case $^4$He is either the projectile or the target, otherwise $E_{cut} = \SI{115}{MeV/u}$.
	\begin{equation}
		f_\textup{cut1} = \frac{1}{1 + e^\frac{-E+E_\textup{cut}}{d}} 
		\label{Eq:f_cut_1}
	\end{equation}
	\begin{equation}
		f_\textup{cut2} = \frac{1}{1 + e^\frac{E-E_\textup{cut}}{d}}
		\label{Eq:f_cut_2}
	\end{equation}
	with $d=\SI{1}{MeV/u}$. It has been noticed that $f_\textup{cut1}$ and $f_\textup{cut2}$ were inverted in Ref.\cite{Sihver2014HK}.
	The values of the Tripathi parameterization are renormalized so that they match the ``Kurotama'' value at $E_\textup{cut}$.
	The Hybrid-Kurotama parameterization has been implemented within this work in the same way it is in PHITS, i.e. using Tripathi99 (section~\ref{sec:Tri99}) for $\sigma_\textup{Trip}$ in the case of ``light'' nucleus-nucleus systems ($A \le 4$ for at least either the projectile or the target) and Tripathi96 otherwise (section~\ref{sec:Tri96}).
	
	\subsection{Projectile-target asymmetry issue}
	Due to the neutron-excess term, none of the parameterizations give the same result for target-projectile exchange, unless they are characterized by the same $A-2Z$. This is a non-physical result, as reaction cross-sections should not depend on the reference system.
	The heavier the target nucleus, the stronger the effect of the neutron-excess term. The heavier nucleus always plays the role of the target in the models, for how they are implemented within this work.
	For high energies, the removal of the neutron-excess parameter tends to fit the experimental data better, but it is not valid for intermediate ($>\SI{200}{MeV/u}$ and $<\SI{1}{GeV/u}$) and especially for low ($\le \SI{200}{MeV/u}$) energies.
	
	
	\section{Comparison of the parameterizations with the experimental data collection}
	\label{sec:results}
\begin{figure}[htb]
		\centering
		\subfloat[][]
		{\includegraphics[width=.55\textwidth]{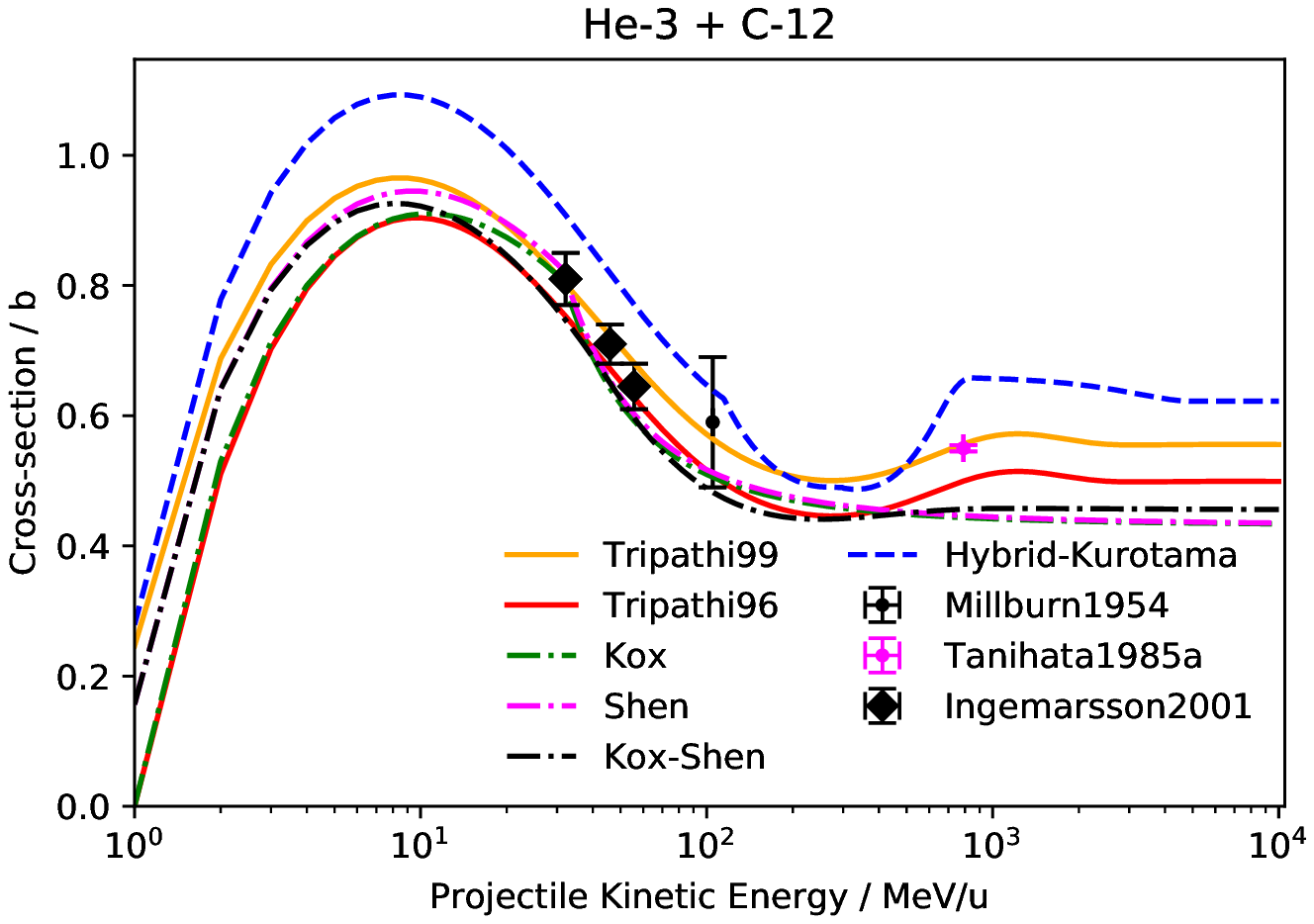}}
		\subfloat[][]
		{\includegraphics[width=.55\textwidth]{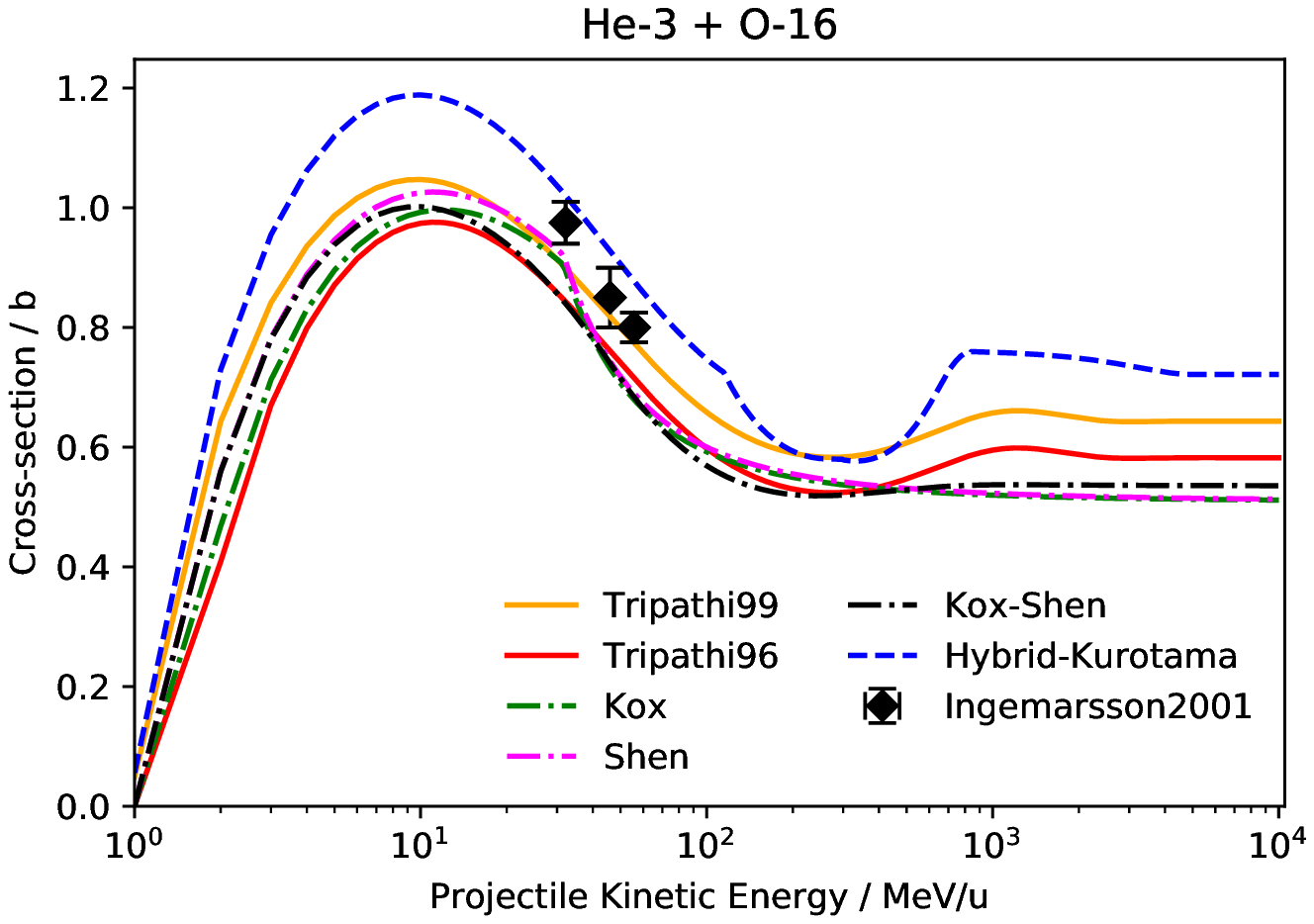}}
		\caption{Comparison between parameterization results and cross-section data for the following systems: $^3$He + $^{12}$C, $^{16}$O. Data in panel (a) are from references~\cite{Millburn1954,Tanihata1985_1,Ingemarsson2001}, in panel (b) from reference~\cite{Ingemarsson2001}.}
		\label{fig:He3}
	\end{figure}

\begin{figure}[htb]
		\centering
		\subfloat[][]
		{\includegraphics[width=.55\textwidth]{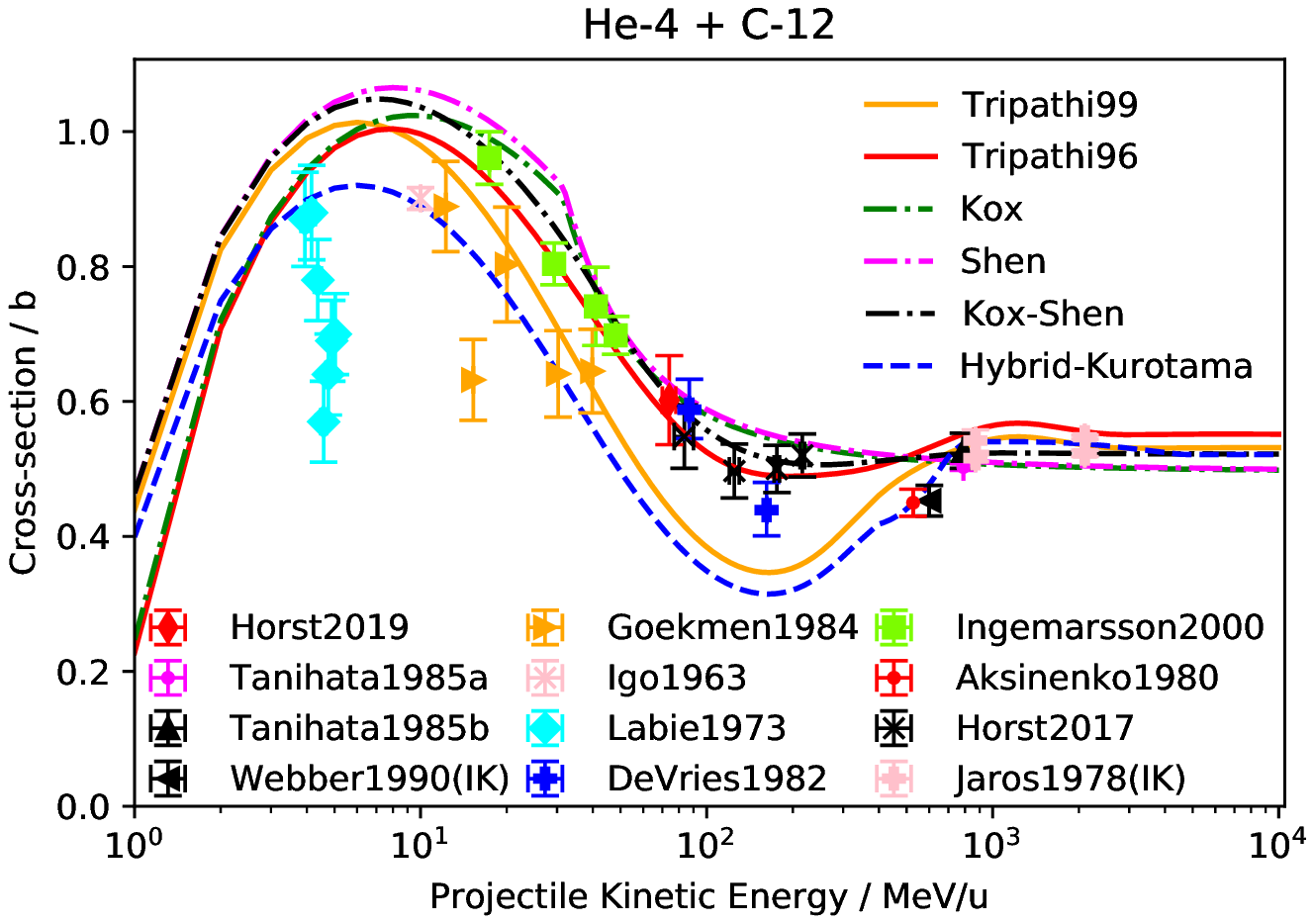}} \\
		\subfloat[][]
		{\includegraphics[width=.55\textwidth]{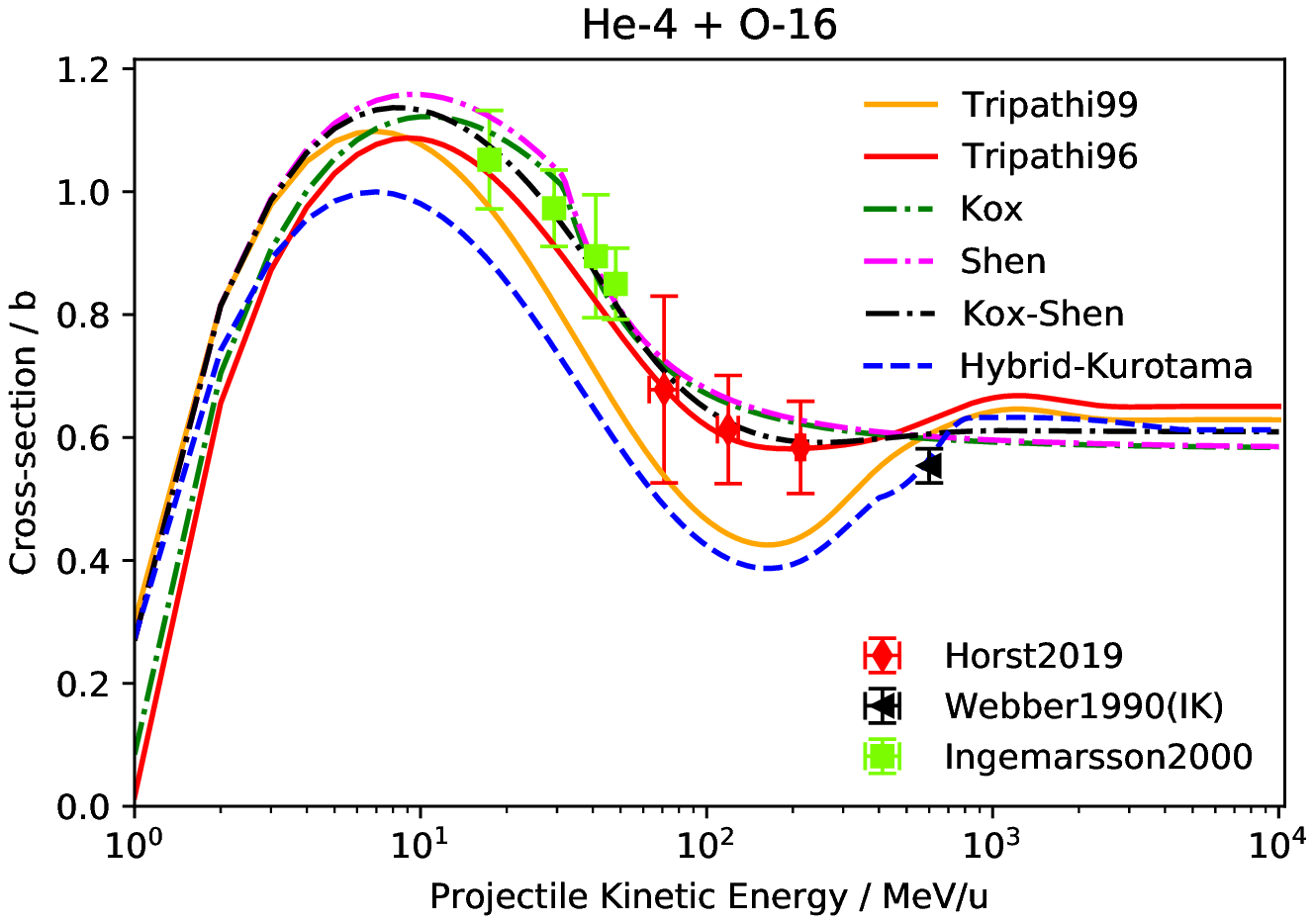}}
		\subfloat[][]
		{\includegraphics[width=.55\textwidth]{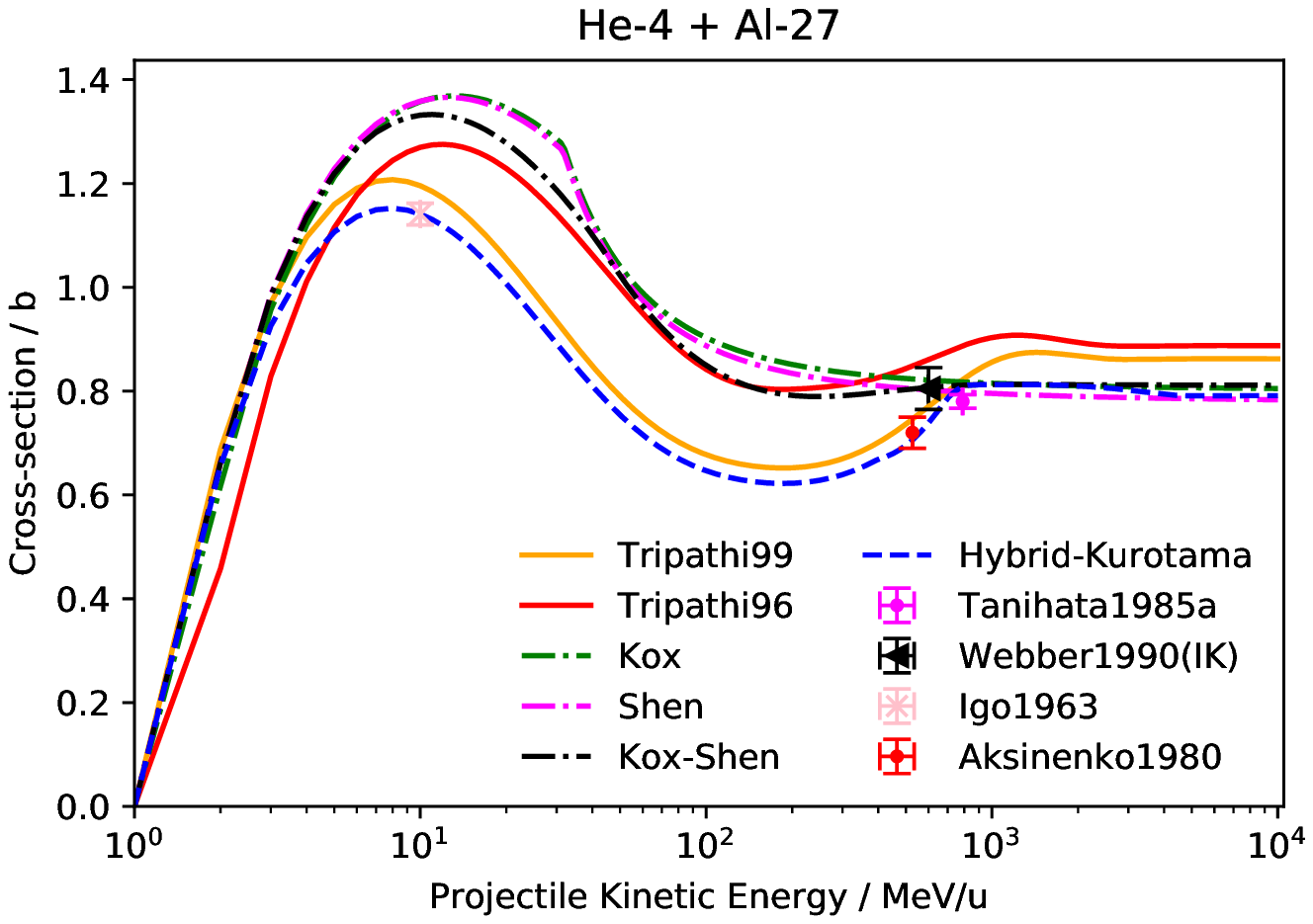}} \\
		\subfloat[][]
		{\includegraphics[width=.55\textwidth]{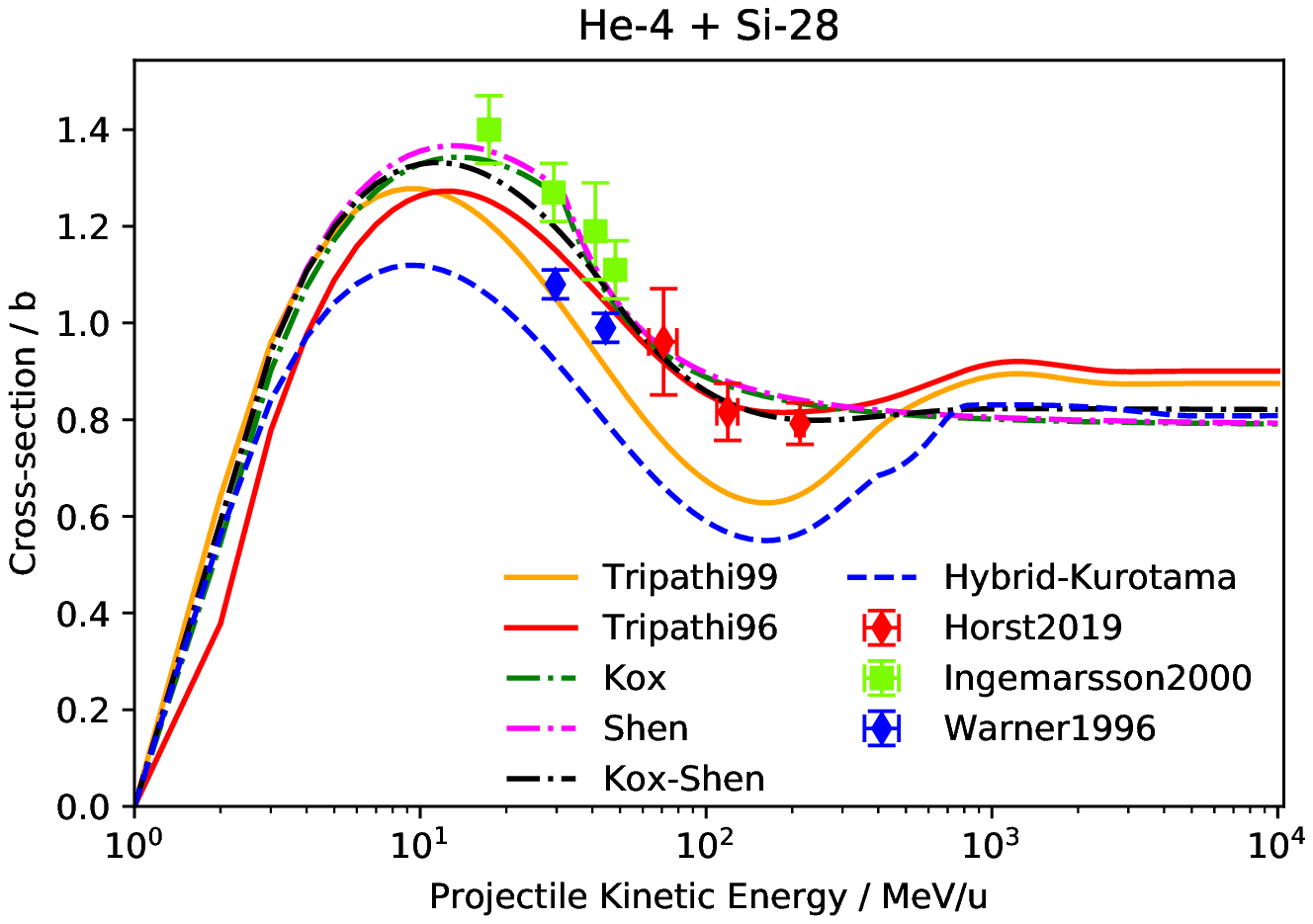}}
		\caption{Comparison between parameterization results and cross-section data for the following systems: $^4$He + $^{12}$C, $^{16}$O, $^{27}$Al, $^{28}$Si. IK stands for Inverse Kinematic data. To be noted that Tripathi96 is not the original model \cite{Tripathi1996}, but optimizations from Horst \textit{et al.} \cite{Horst2019} for $^4$He projectile are included. Data in panel (a) are from references~\cite{Horst2019,Tanihata1985_1,Tanihata1985_2,Webber1990,Goekmen1984,Igo1963,Labie1973,DeVries1982,Ingemarsson2000,Aksinenko1980,Horst2017,Jaros1978}, in panel (b) from references~\cite{Horst2019,Webber1990,Ingemarsson2000}, in panel (c) from references~\cite{Tanihata1985_1,Webber1990,Igo1963,Aksinenko1980}, in panel (d) from references~\cite{Horst2019,Ingemarsson2000,Warner1996}.}
		\label{fig:He4}
	\end{figure}

\begin{figure}[htb]
		\centering
		\subfloat[][]
		{\includegraphics[width=.55\textwidth]{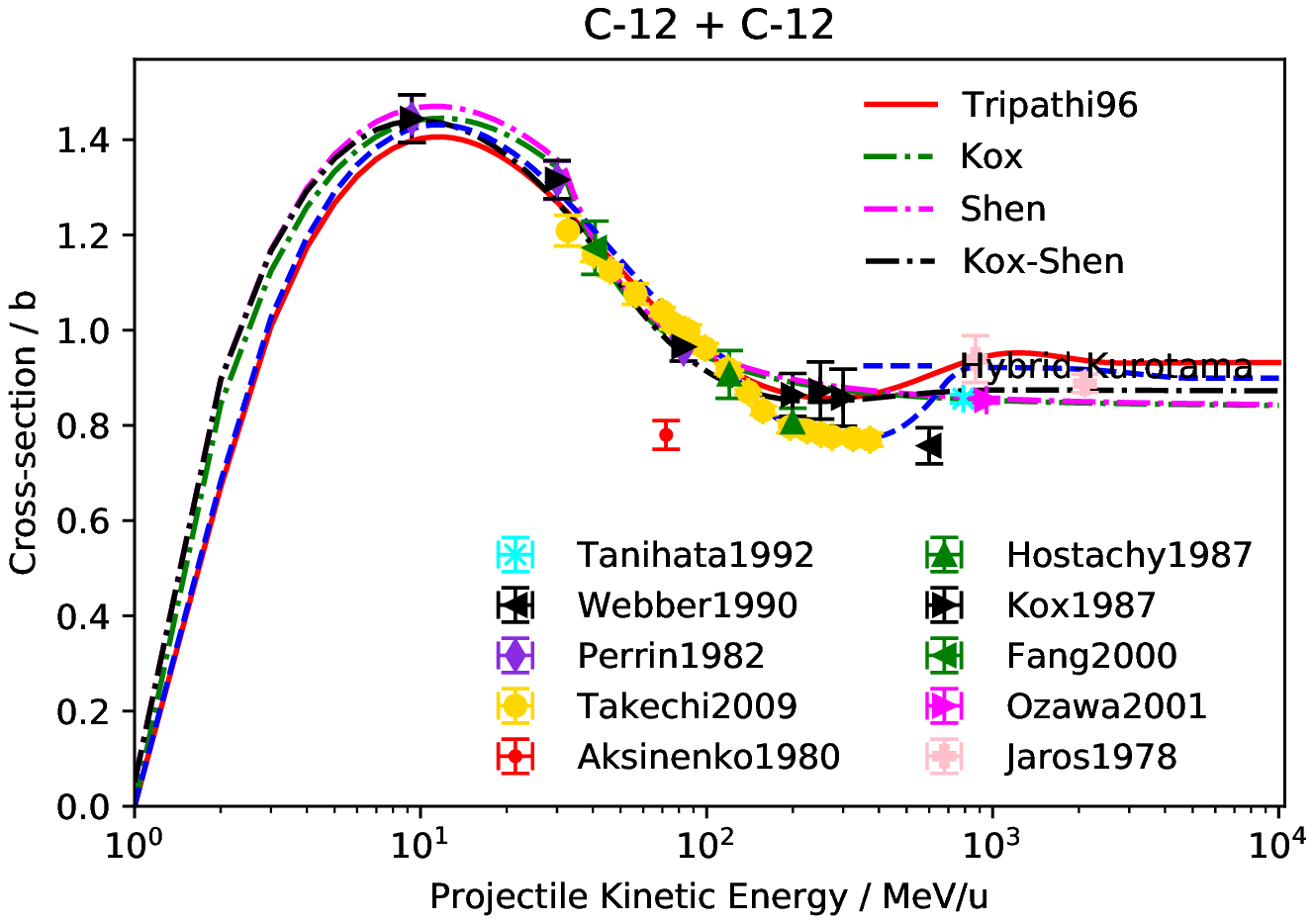}}
		\subfloat[][]
		{\includegraphics[width=.55\textwidth]{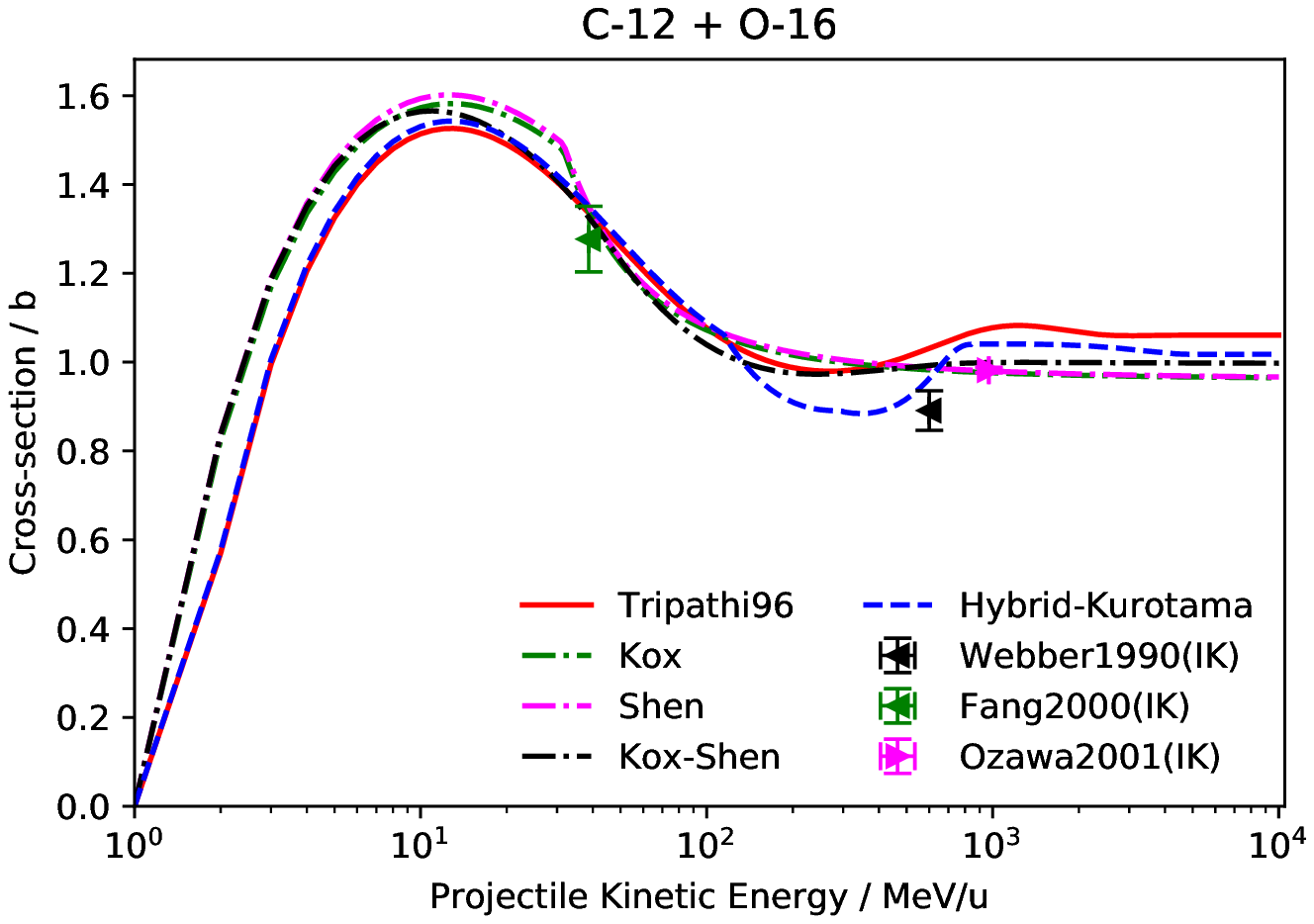}} \\
		\subfloat[][]
		{\includegraphics[width=.55\textwidth]{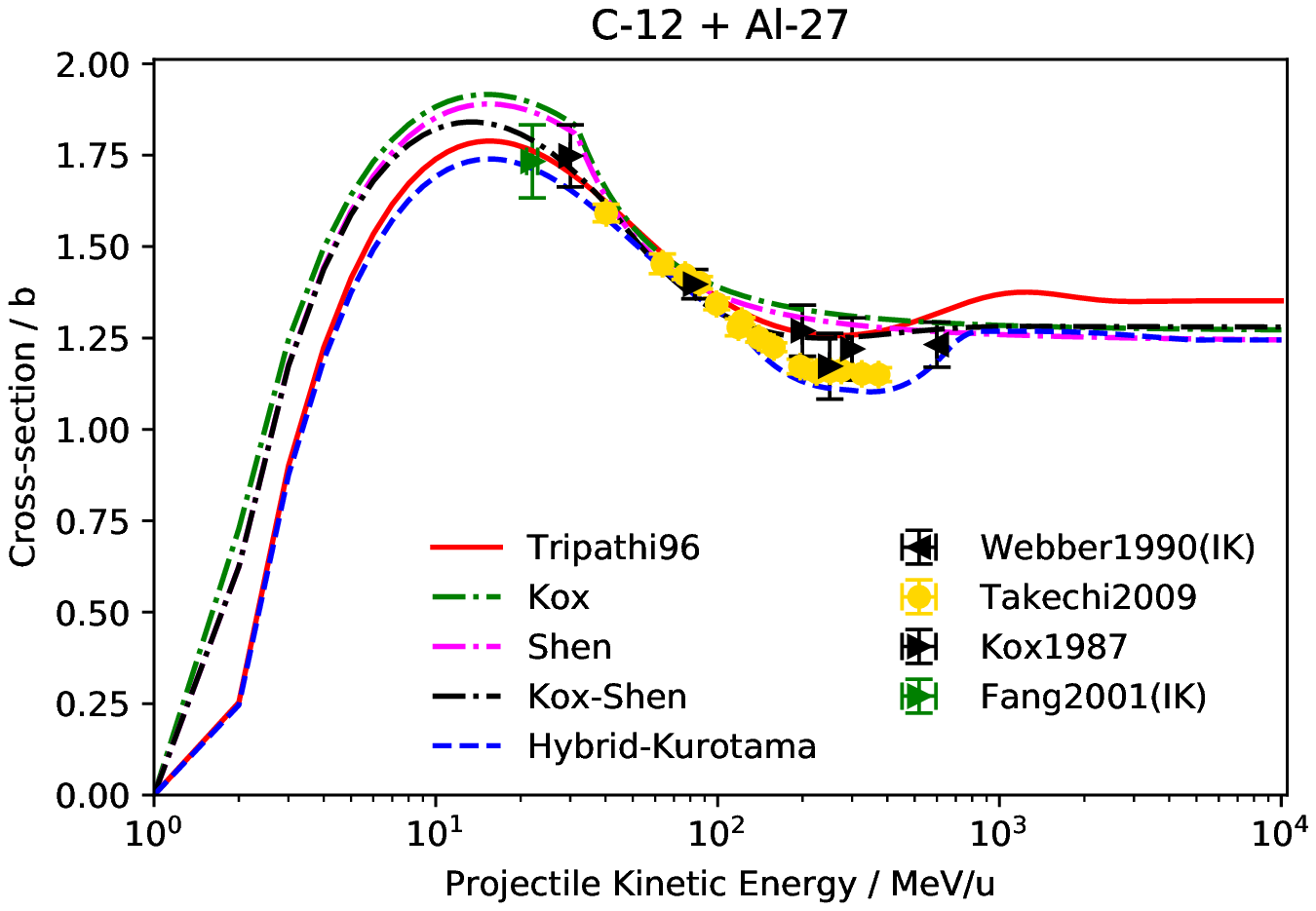}}
		\subfloat[][]
		{\includegraphics[width=.55\textwidth]{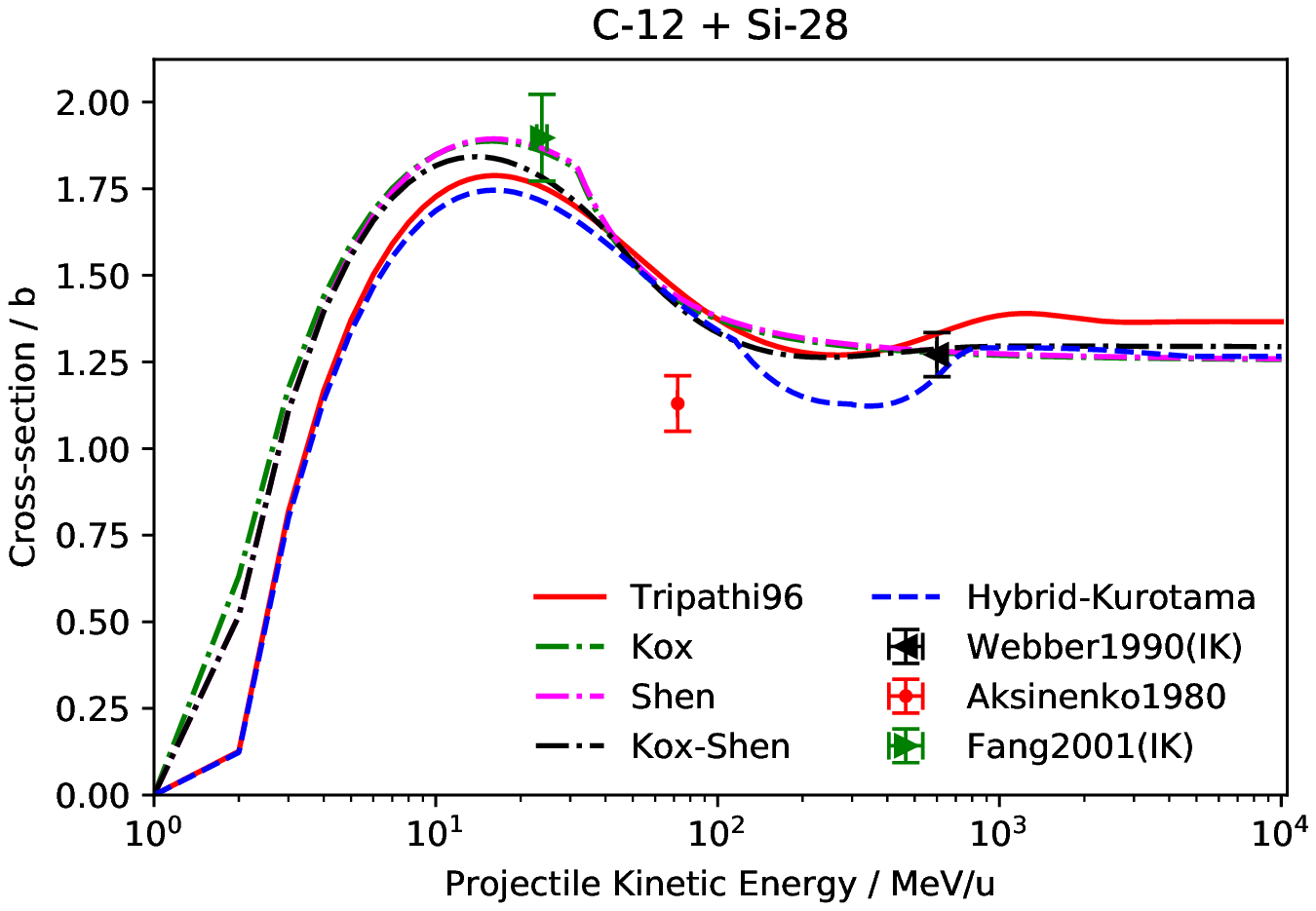}}
		\caption{Comparison between parameterization results and cross-section data for the following systems: $^{12}$C + $^{12}$C, $^{16}$O, $^{27}$Al, $^{28}$Si. IK stands for Inverse Kinematic data. Data in panel (a) are from references~\cite{Tanihata1992,Webber1990,Perrin1982,Takechi2009,Aksinenko1980,Hostachy1987,Kox1987,Fang2000,Ozawa2001,Jaros1978}, in panel (b) from references~\cite{Webber1990,Fang2000,Ozawa2001}, in panel (c) from references~\cite{Webber1990,Takechi2009,Kox1987,Fang2001}, in panel (d) from references~\cite{Webber1990,Aksinenko1980,Fang2001}.}
		\label{fig:C12}
	\end{figure}

\begin{figure}[htb]
		\centering
		{\includegraphics[width=.55\textwidth]{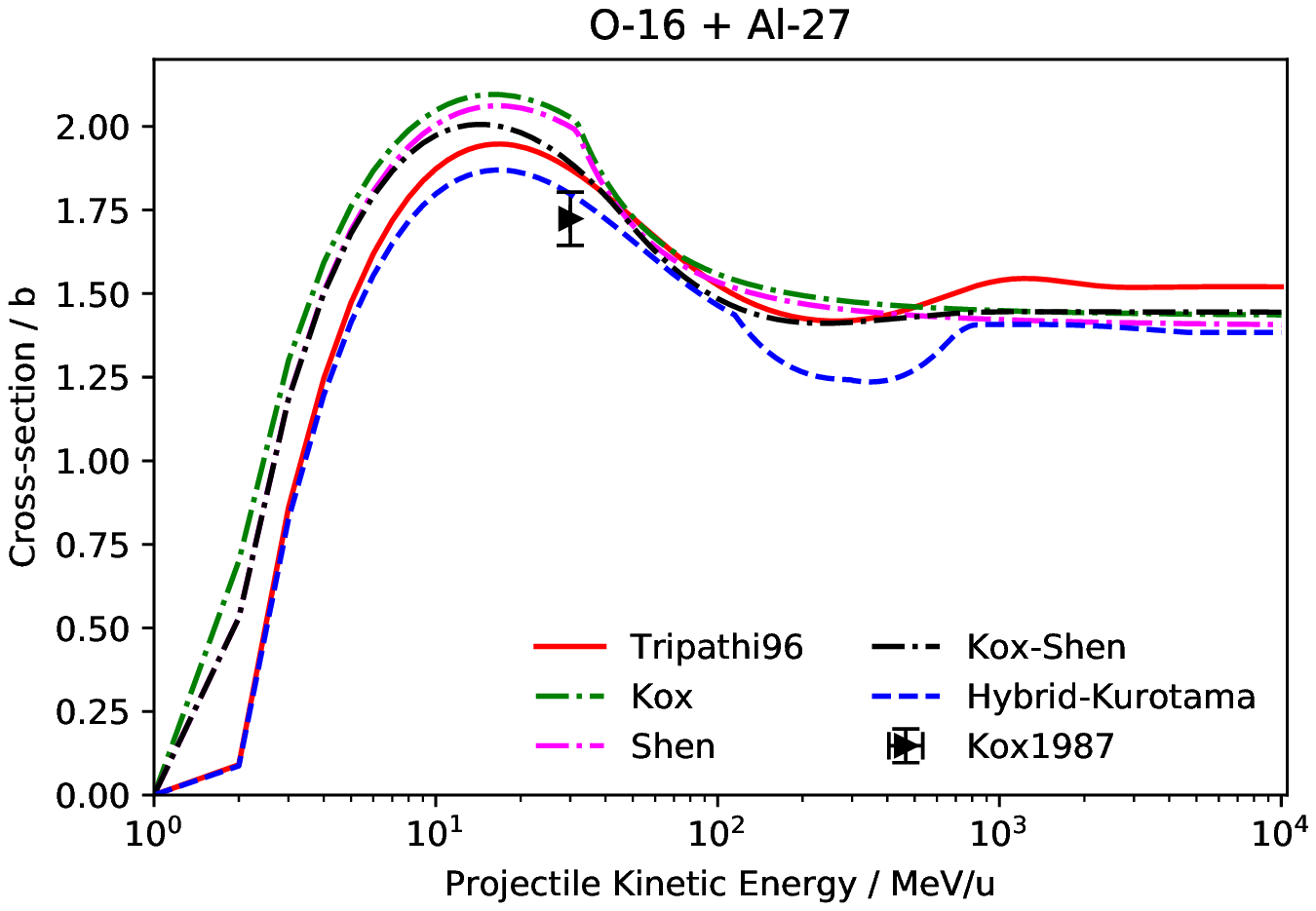}}
		\caption{Comparison between parameterization results and cross-section data for the following system: $^{16}$O + $^{27}$Al. The data point is from reference~\cite{Kox1987}.}
		\label{fig:O16}
	\end{figure}

\begin{figure}[htb]
		\centering
		\subfloat[][]
		{\includegraphics[width=.55\textwidth]{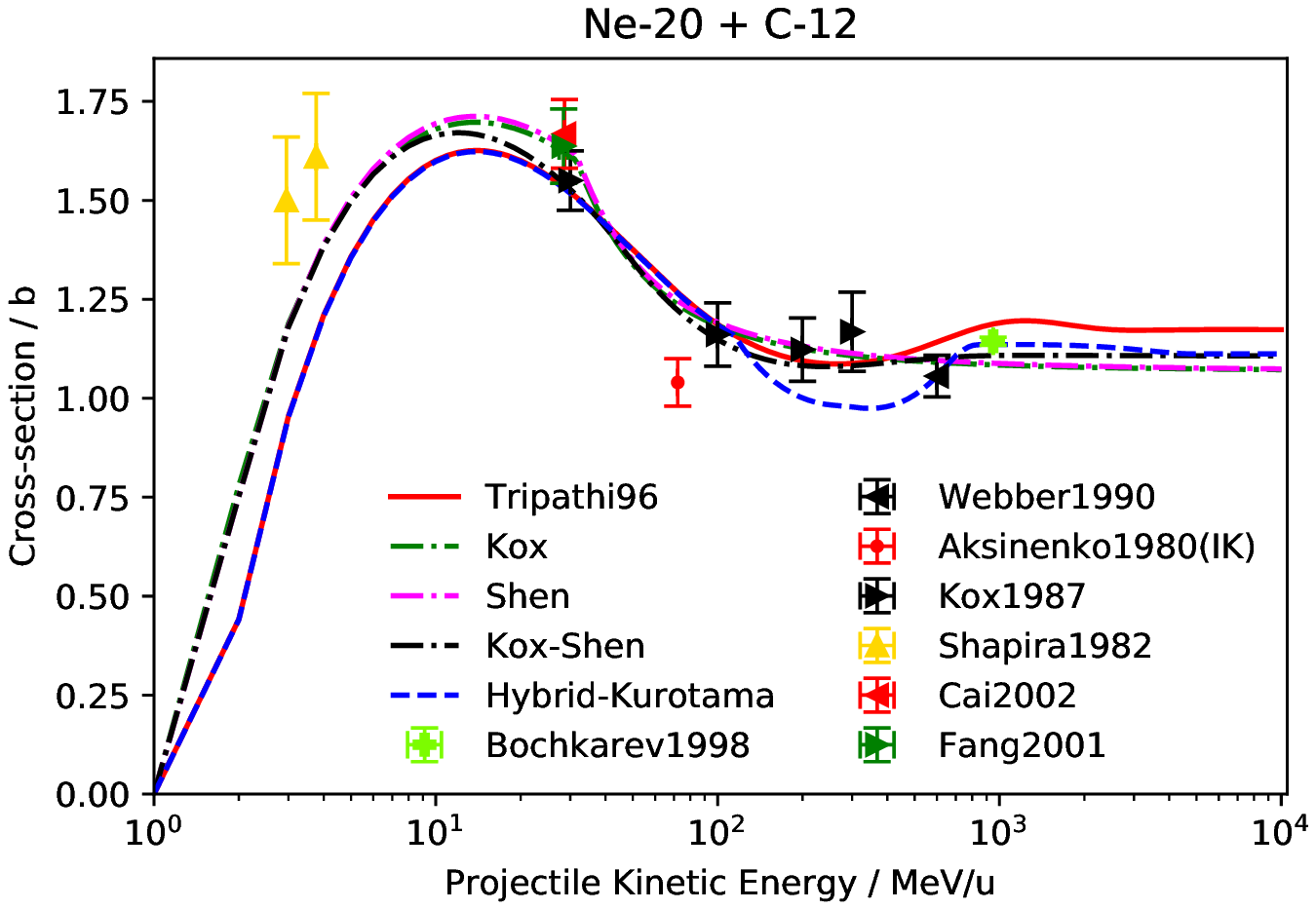}} 
		\subfloat[][]
		{\includegraphics[width=.55\textwidth]{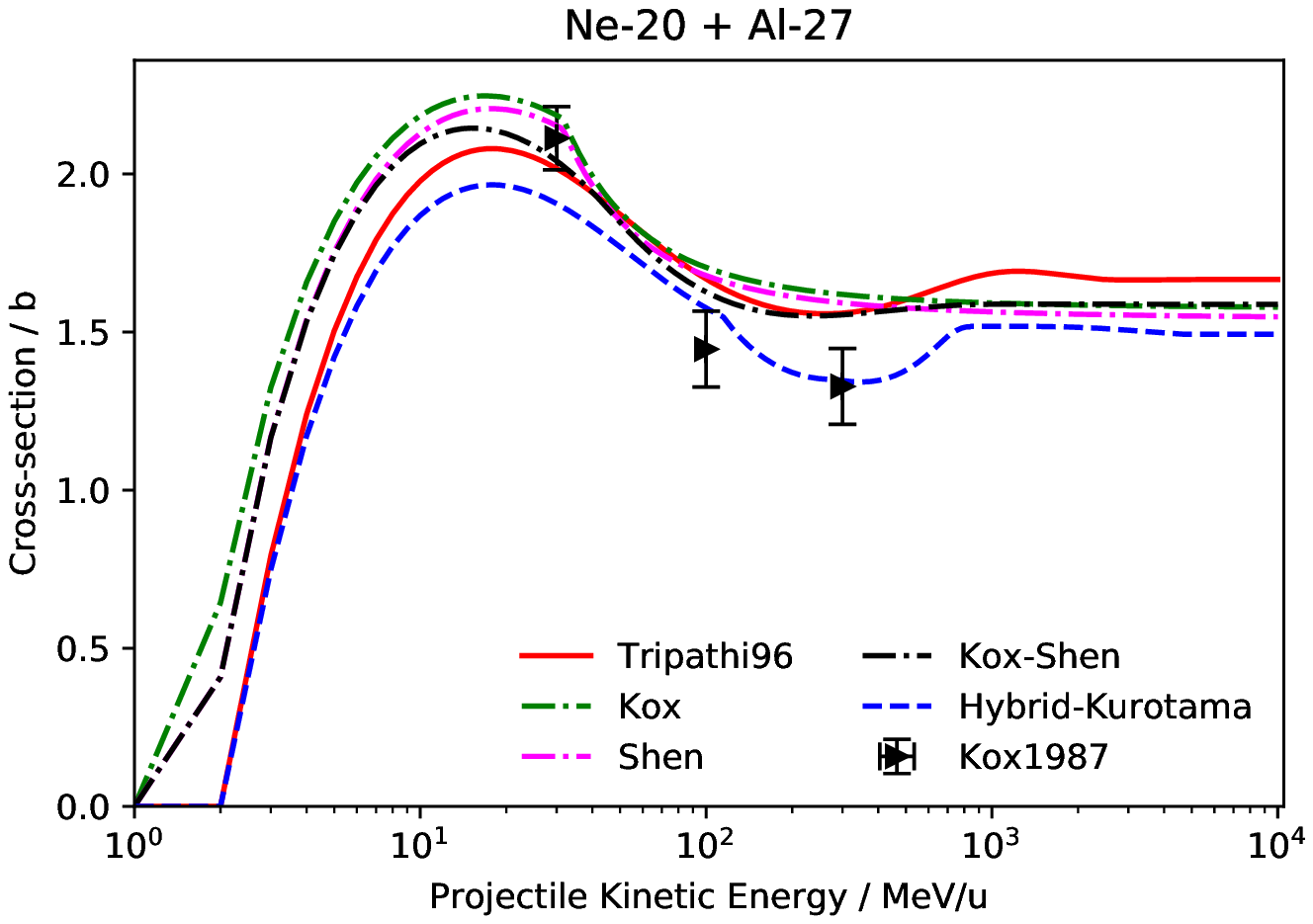}}
		\caption{Comparison between parameterization results and cross-section data for the following systems: $^{20}$Ne + $^{12}$C, $^{27}$Al. IK stands for Inverse Kinematic data. Data in panel (a) are from references~\cite{Bochkarev1998,Webber1990,Aksinenko1980,Kox1987,Shapira1982,Cai2002,Fang2001}, in panel (b) from reference~\cite{Kox1987}.}
		\label{fig:Ne20}
	\end{figure}

\begin{figure}[htb]
		\centering
		{\includegraphics[width=.55\textwidth]{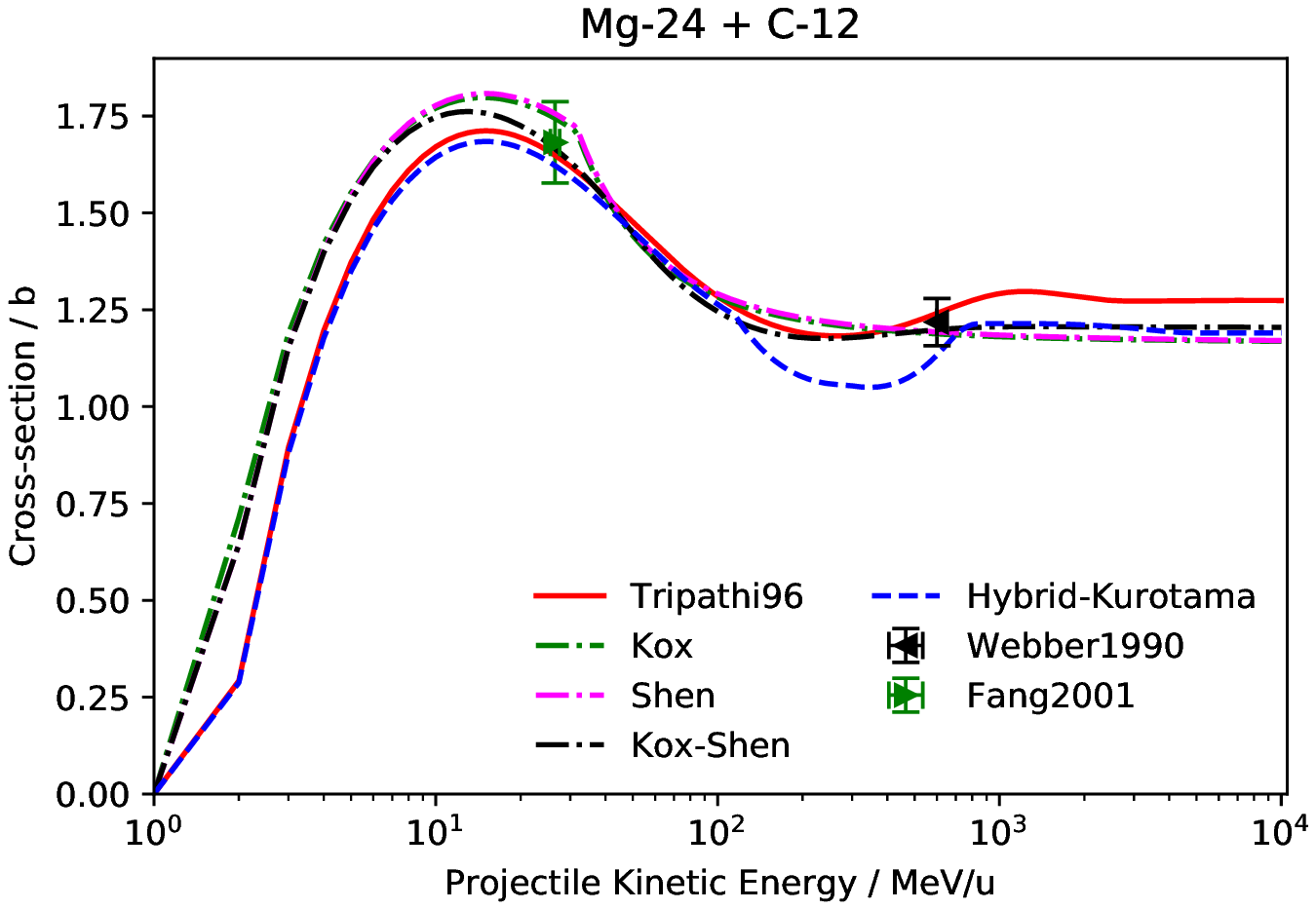}} 
		\caption{Comparison between parameterization results and cross-section data for the following system: $^{24}$Mg + $^{12}$C. Data are from references~\cite{Webber1990,Fang2001}.
		}
		\label{fig:Mg24}
	\end{figure}

\begin{figure}[htb]
		\centering 
		{\includegraphics[width=.55\textwidth]{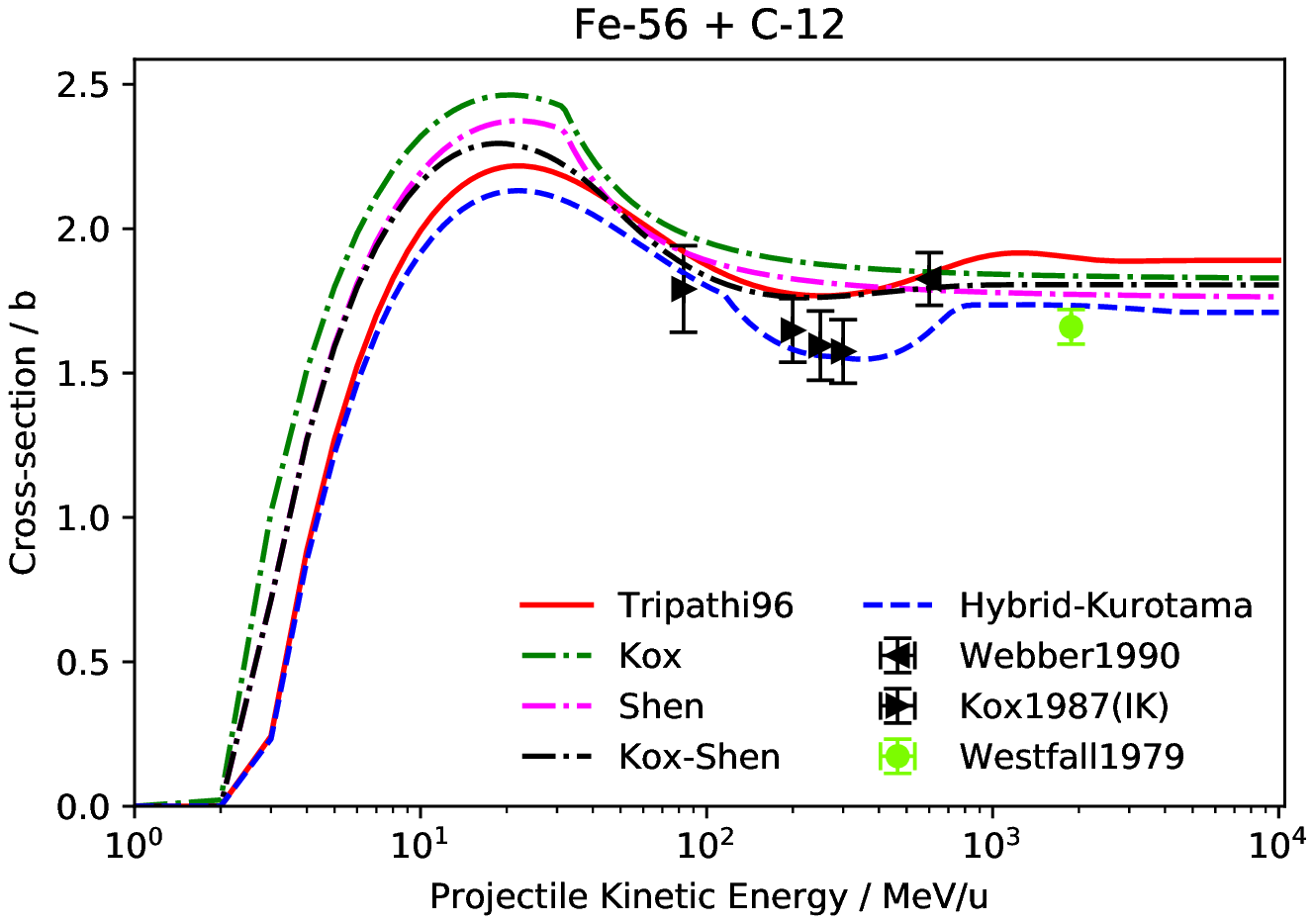}} 
		\caption{Comparison between parameterization results and cross-section data for the following system: $^{56}$Fe + $^{12}$C. IK stands for Inverse Kinematic data. Data are from references~\cite{Webber1990,Kox1987,Westfall1979}.}
		\label{fig:Fe56}
	\end{figure}

	The primary systems of interest for space radiation and heavy-ion therapy applications are: 
	\begin{equation}
		^3\textup{He} + ^{12}\textup{C}, \, ^{16}\textup{O}
		\label{Eq:systems_therapy}
	\end{equation}
	and
	\begin{equation}
		^4\textup{He}, \, ^{12}\textup{C}, \, ^{16}\textup{O}, \, ^{20}\textup{Ne}, \, ^{24}\textup{Mg}, \, ^{28}\textup{Si}, \, ^{56}\textup{Fe} + ^{12}\textup{C}, \, ^{16}\textup{O}, \, ^{27}\textup{Al}, \, ^{28}\textup{Si}.
		\label{Eq:systems_therapyandspace}
	\end{equation}
	In concern with projectiles, $^3$He has recently shown to be an interesting candidate for heavy-ion therapy \cite{Horst2021}. $^4$He and $^{12}$C are important because of their contribution to the dose in space due to GCR~\cite{Norbury2020} and also because of their importance for particle therapy purposes~\cite{Gruen2015,Norbury2020}. $^{16}$O, $^{20}$Ne, $^{24}$Mg and $^{56}$Fe have been chosen because of their relative importance in the dose equivalent in free space and behind thin shields. 
	As targets, $^{12}$C and $^{16}$O have been chosen because they are among the main components of the human body. $^{27}$Al is the most important structural material spacecrafts are made of. $^{28}$Si is the main component of electronic devices and regoliths. Regoliths of planets and the Moon can be in fact, exploited for \textit{in situ} shielding.
	In figures~\ref{fig:He3} to~\ref{fig:Fe56}, available cross-section data extracted from the database for a few projectile-target systems are reported alongside the different models.
	Also inverse kinematics data are reported in the plots. 
	For this reason and since the reimplementation of each model uses the lightest nucleus as projectile, some plots are not reported because they would be identical to others. E.g. $^{28}$Si + $^{12}$C would be identical to $^{12}$C + $^{28}$Si.
	Only mass-changing and reaction cross-sections are shown (see section~\ref{sec:cc_mc}). 
	
	Such plots can be directly generated on the web application developed as part of the work (see section~\ref{sec:webpage}).
	
	In Figure~\ref{fig:He3}, total reaction cross-section data of $^3$He projectiles impinging on $^{12}$C and $^{16}$O are plotted alongside the parameterizations. The error bars of all data are both statistical and systematic. Only for the first system a comparison between models and data is possible both at low and high energies. 
	Charge-changing cross-section data are plotted as well (Millburn1954), since for $^3$He $\sigma_{cc} = \sigma_R$ (see section~\ref{sec:cc_mc}).
	For $^3$He + $^{12}$C, all the models seem to reproduce the data well, Tripathi99 being the best fit both for low and high energies. Hybrid-Kurotama overestimates the high-energy data point and consequently (since Hybrid-Kurotama uses Tripathi scaled to the Kurotama predictions at $E_\textup{cut}$) also the low-energy dataset. Tripathi96, Kox, Shen and Kox-Shen underestimate the high-energy point. However, no intermediate energy data were measured for such targets. 
	The Hybrid-Kurotama predictions also slightly overestimate the $^3$He + $^{16}$O dataset, while Tripathi96, Kox, Shen and Kox-Shen underestimate it. Tripathi99 seems to be the best fit for this system as well. 
	Nevertheless, for any further model validation in the scope of $^3$He-ion therapy, also a single measurement point at $\SI{200}{MeV/u}$ on H$_\textup{2}$O targets is available \cite{Horst2021} (see Figure~\ref{fig:H2Odata}). This data point was measured for intermediate energies, which are the most relevant to therapy application. As it can be seen in Figure~\ref{fig:H2Odata}, the Kox-Shen model, which predicts the lowest cross-section values for intermediate energies, reproduces the measurement well, while Tripathi99 would overestimate it. Therefore, the Kox-Shen model is recommended by the authors for $^3$He-ion therapy applications.
	
	In Figure~\ref{fig:He4}, total reaction cross-section data of $^4$He projectiles impinging on the different targets are plotted alongside the parameterizations. Error bars of the Webber1990 and Igo1963 datasets are not specified to be only statistical or systematic as well. Labie1973 and DeVries1982 error bars are only statistical. The rest are both statistical and systematic. 
	It is hard to comment on the $^4$He + $^{12}$C system, since some datasets are not compatible with each other. The Labie1973 data collection is likely not optimal for comparison. The reason is that the data do not contain any contributions from compound elastic scattering, which have a resonance in this energy region. 
	The fluctuations of the G\"okmen1984 dataset are non-physical. In addition, the data are not compatible with the Ingemarsson2000 dataset. Since Ingemarsson data are more recent, they are considered to be more reliable. Therefore, the models fitting Ingemarsson results are to be considered more reliable than the models fitting G\"okmen data, for instance. Since the Igo1963 data point is old and lower than Ingemarsson2000, it is believed that it underestimates the real cross-section value. The same considerations apply to the Warner1996 data (see $^4$He + $^{28}$Si).
	The Horst2019, Horst2017 and DeVries1982 data are compatible within error bars. 
	It must be noted that Aksinenko1980 data (see $^4$He + $^{28}$Si) are also believed to underestimate the real cross-section values, as explained later in this section.
	For all four of the systems, enough data points are available to check if the models fit the data well for low, mid and high energies.
	Regarding $^4$He nuclei impinging on any target, all the models seem to be in agreement with the data points. Nevertheless, Tripathi99 and as a consequence, Hybrid-Kurotama show a tendency to underestimate the data at low and intermediate energies. The only data points that are well fitted by Hybrid-Kurotama at low energies are indeed Igo1963 and G\"okmen1984, which are believed to underestimate the real cross-sections themselves for the reasons explained above. The model that seems to fit the data best is Tripathi96, thanks to the optimizations recently proposed \cite{Horst2019}.
	Kox, Shen and Kox-Shen seem to be the models fitting the data best at high energies (see $^4$He + $^{12}$C).
	
	In Figure~\ref{fig:C12}, data for $^{12}$C on different targets are shown alongside the parameterization results. Error bar types of the Webber1990 and Takechi2009 datasets are not specified to be only statistical or systematic too. Hostachy1987 error bars are only statistical. The rest are both statistical and systematic. 
	In concern with the $^{12}$C + $^{12}$C system, the only data point that looks incompatible with the others in the low-energy range ($\sim \SI{70}{MeV/u}$) is Aksinenko1980. It is in fact systematically smaller than the data from Takechi2009, Kox1987 and Hostachy1987. It is the oldest data point in this energy range. Therefore, it is reasonable to believe that the results of Ref.~\cite{Aksinenko1980} are not reliable. For this reason, it is not surprising that none of the models fit the Aksinenko1980 data point for the $^{12}$C + $^{28}$Si system. In the mid-energy range ($\sim \SI{250}{MeV/u}$) though, the Kox1987 data points are higher than both Takechi2009 and Hostachy1987 data. Therefore, it is believed that in this energy region the last two datasets are the most precise.
	Coming to $^{12}$C + $^{28}$Si, for all four of the systems, enough data points are available to check if the models fit the data well for low, mid and high energies.
	All the models seem to be in agreement with the data, but Kox, Shen and Kox-Shen cannot reproduce the oscillation in the data (dip in the mid-energy region), which is parameterized by the cosine term in $C_E$ in the Tripathi model, and consequently in the Hybrid-Kurotama model. 
	Kox, Shen and Kox-Shen seem to be the models fitting the data best at high energies (see $^{12}$C + $^{12}$C and $^{12}$C + $^{16}$O).
	
	Concerning $^{16}$O projectiles on targets of interest for radiation protection in space, from Figure~\ref{fig:O16} it is clear that data are missing. There is no data either for $^{16}$O or $^{28}$Si targets. There is only one data point for $^{27}$Al. $^{16}$O + $^{12}$C is not reported since it would be the same plot as $^{12}$C + $^{16}$O\footnote{For every model the lightest nucleus is used as the projectile. So for both cases the models would be identical. Since  inverse kinematic data is included, both figures ($^{12}$C + $^{16}$O and $^{16}$O + $^{12}$C) would show the same data.}.
	Uncertainties are both statistical and systematic. There are not enough statistics to compare the model predictions with the data.
	
	In Figure~\ref{fig:Ne20}, data for $^{20}$Ne projectiles are shown. Bochkarev1998 error bars are only statistical, while Webber1990 error bar type is not specified. The rest are both statistical and systematic. 
	Concerning $^{20}$Ne + $^{12}$C, data for all energy ranges were measured, even within the Coulomb barrier. Also in this case, the Aksinenko1980 data point is lower than the data points from the other sets. It is interesting to notice that for mid energies the Kox1987 values are higher than Webber1990, which is different behavior to other data discussed below. Three different measurements were performed around $\SI{30}{MeV/u}$. They are compatible with each other within error bars.
	All the models seem to reproduce the data reasonably well, even if all of them predict lower values than the Shapira1982 data points.
	For the case of $^{20}$Ne + $^{27}$Al, there is not enough statistics to compare the model predictions with the data since all data were collected within the same measurement campaign.
	No experimental data were measured for $^{20}$Ne + $^{16}$O or $^{20}$Ne + $^{28}$Si.
	
	Also for the case of $^{24}$Mg, not many experimental data were measured. Only for the case of $^{24}$Mg + $^{12}$C, two data were measured (see Figure~\ref{fig:Mg24}). The Webber1990 error bar type is not specified and Fang2001 are both statistical and systematic. There is not enough statistics to compare the model predictions with the data.
	
	No experimental data were measured for $^{28}$Si + $^{27}$Al or $^{28}$Si + $^{28}$Si.
	
	Surprisingly, no data for $^{56}$Fe + $^{16}$O, $^{27}$Al, $^{28}$Si are available (see Figure~\ref{fig:Fe56}). For what concerns $^{56}$Fe + $^{12}$C, Webber1990 error bar type is not specified, the rest are both statistical and systematic. The Kox1987 values are lower than Webber1990. It is difficult to say which model reproduces the data best.
	
\begin{figure}[htb]
		\centering 
		\subfloat[][]
		{\includegraphics[width=.55\textwidth]{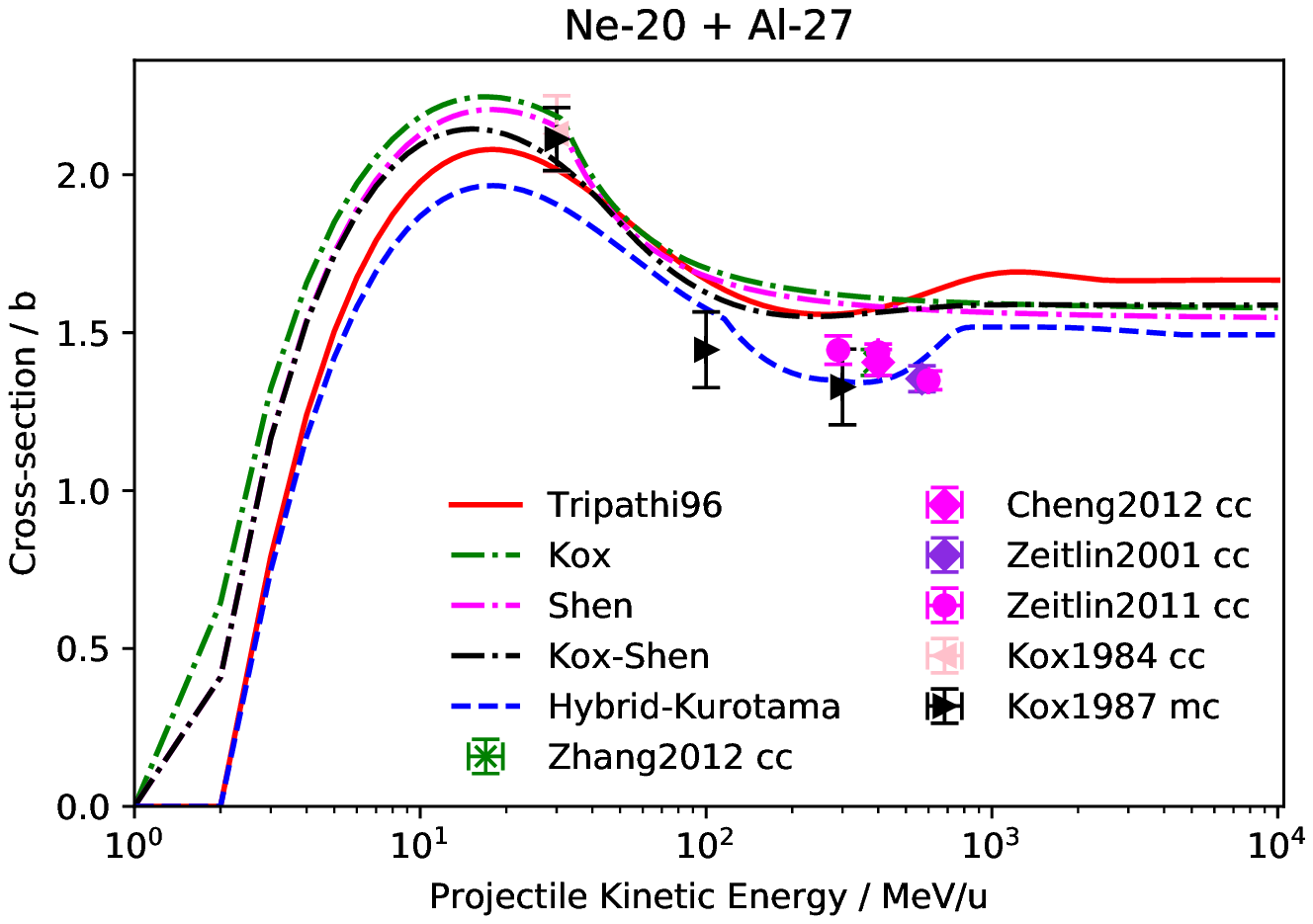}} 
		\subfloat[][]
		{\includegraphics[width=.55\textwidth]{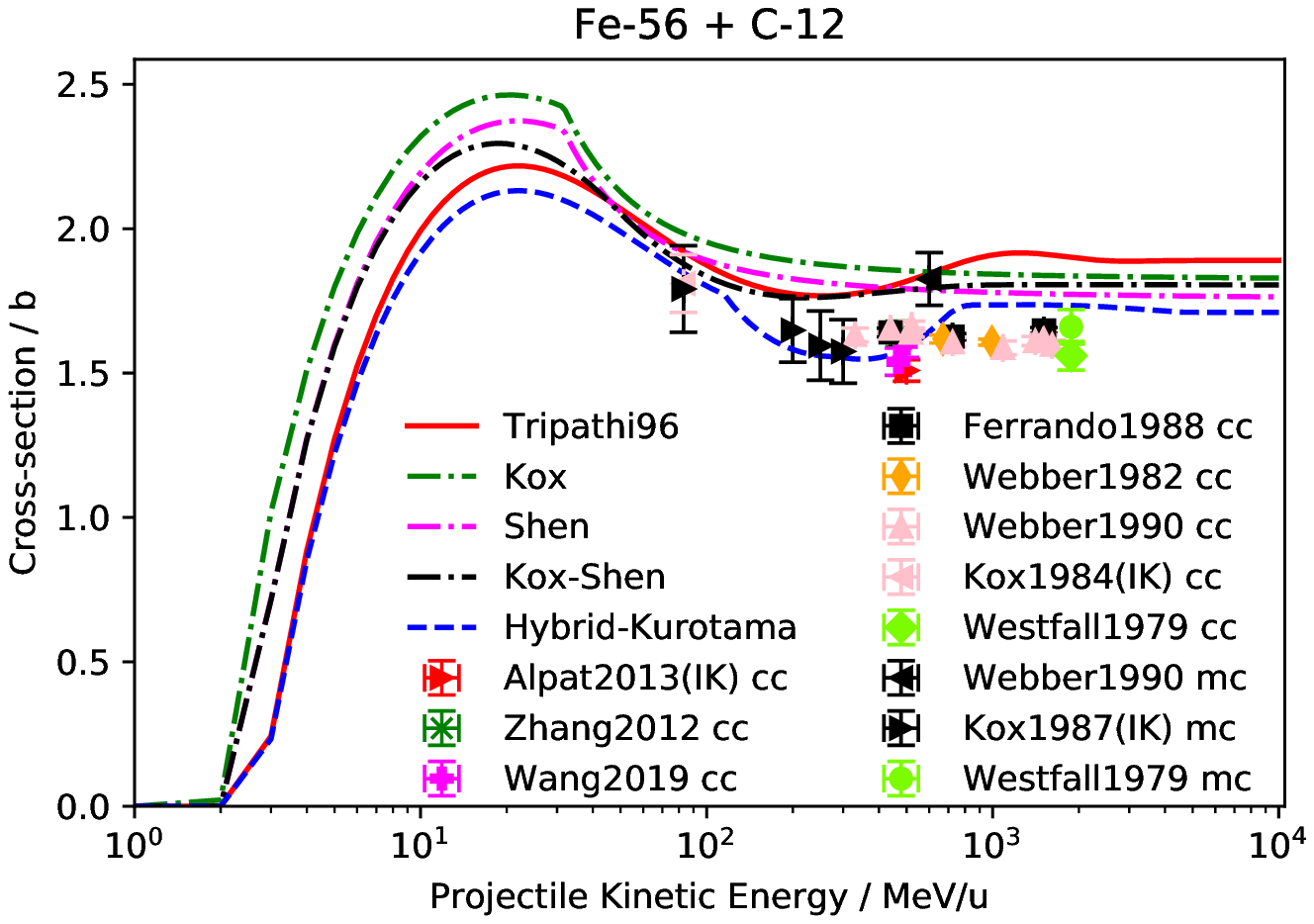}} 
		\caption{Comparison between parameterization results and cross-section data for the following systems: $^{20}$Ne + $^{27}$Al and $^{56}$Fe + $^{12}$C. IK stands for Inverse Kinematic data, ``cc'' for charge-changing and ``mc'' for mass-changing cross-sections. Data in panel (a) are from references~\cite{Zhang2012,Cheng2012,Zeitlin2001,Zeitlin2011,Kox1984,Kox1987}, in panel (b) from reference~\cite{Alpat2013,Zhang2012,Wang2019,Ferrando1988,Webber1982,Webber1990,Kox1984,Westfall1979,Kox1987}.}
		\label{fig:Kox1987}
	\end{figure}

	Some additional evidence about the systematic underestimation of the reaction cross-section data by Kox1987 dataset comes from the comparison with charge-changing cross-section data measured at comparable energies for $^{20}$Ne + $^{27}$Al and $^{56}$Fe + $^{12}$C (see Figure~\ref{fig:Kox1987}). The charge-changing cross-sections are indeed, higher than the Kox1987 data. This is not physical since by definition, reaction cross-sections are supposed to be higher than charge-changing cross-sections.
	
\begin{figure}[htb]
		\centering
		\subfloat[][]
		{\includegraphics[width=.55\textwidth]{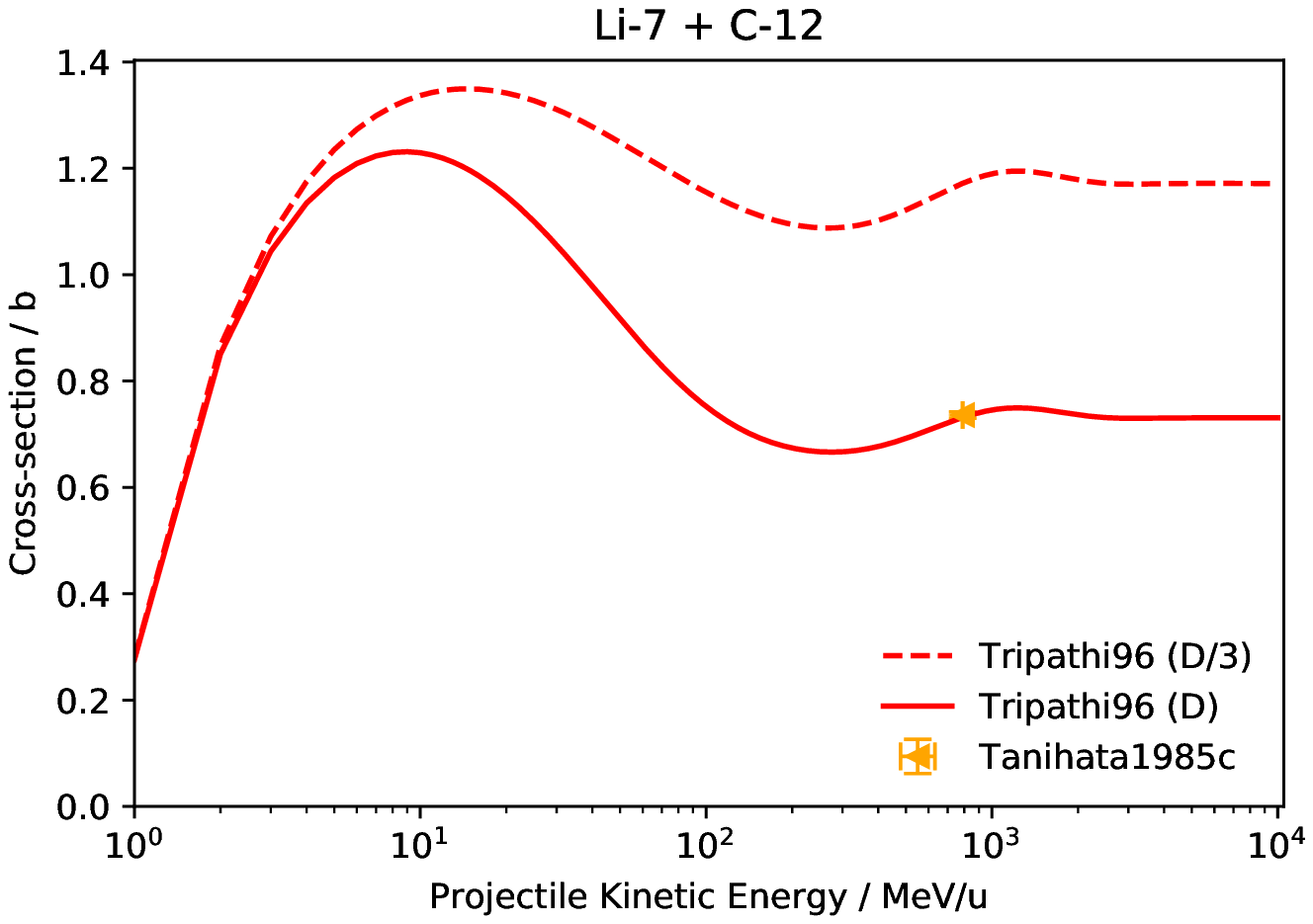}}
		\subfloat[][]
		{\includegraphics[width=.55\textwidth]{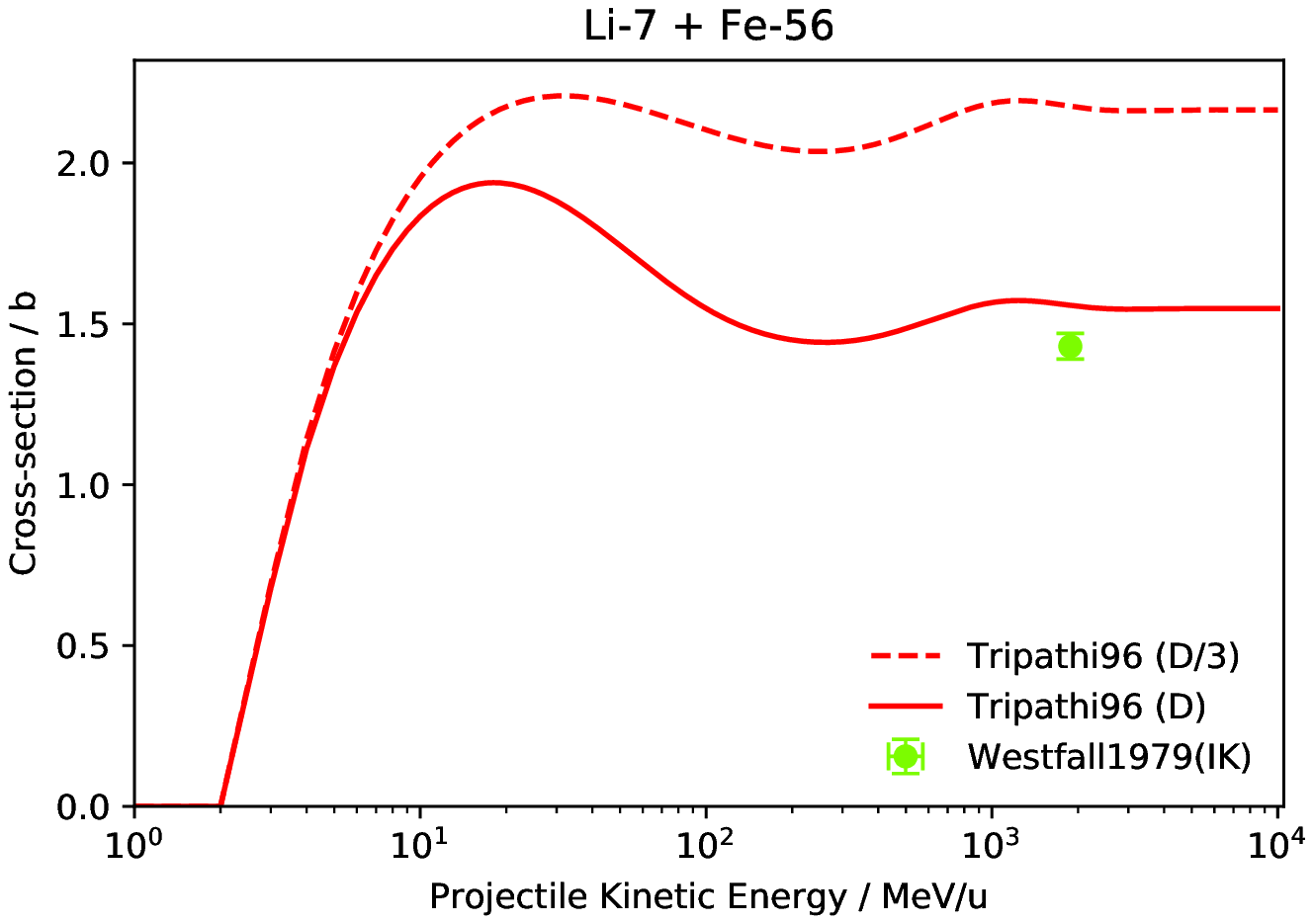}}
		\caption{Tripathi96 computed both using $D$ from Equation~\ref{Eq:D} (choice followed in the present work) and $D/3$ (as recommended in Ref.~\cite{Tripathi1996}) for the following systems: $^7$Li + $^{12}$C, $^{56}$Fe. IK stands for Inverse Kinematic data. Data in panel (a) are from reference~\cite{Tanihata1985_3}, in panel (b) from reference~\cite{Westfall1979}.}
		\label{fig:D_Li}
	\end{figure}

	A few general comments about the parameterizations are added. 
	The literature \cite{Tripathi1996} recommends to use $D/3$ instead of $D$ for lithium projectiles in the Tripathi96 model calculations. However, it has been noticed that using $D$ gives better agreement with the experimental data (see Figure~\ref{fig:D_Li} for a few examples).
	
	As already stated in Ref.~\cite{Shen1989}, the neutron excess parameter plays an important role at low energies ($< \SI{200}{MeV/u}$). However, because of the structure of the parameterizations, it is likewise important at all energy ranges. A careful study has been conducted with the experimental data collected in the database and the Kox-Shen model. It has been confirmed that if the multiplication factor of the neutron excess parameter is set to $\alpha=5$, low-energy data are fitted best, while $\alpha=0$ would fit them better at high energies. The recommended parameter for the model is $\alpha=1$.
	
\begin{figure}[htb]
		\centering 
		\subfloat[][]
		{\includegraphics[width=.55\textwidth]{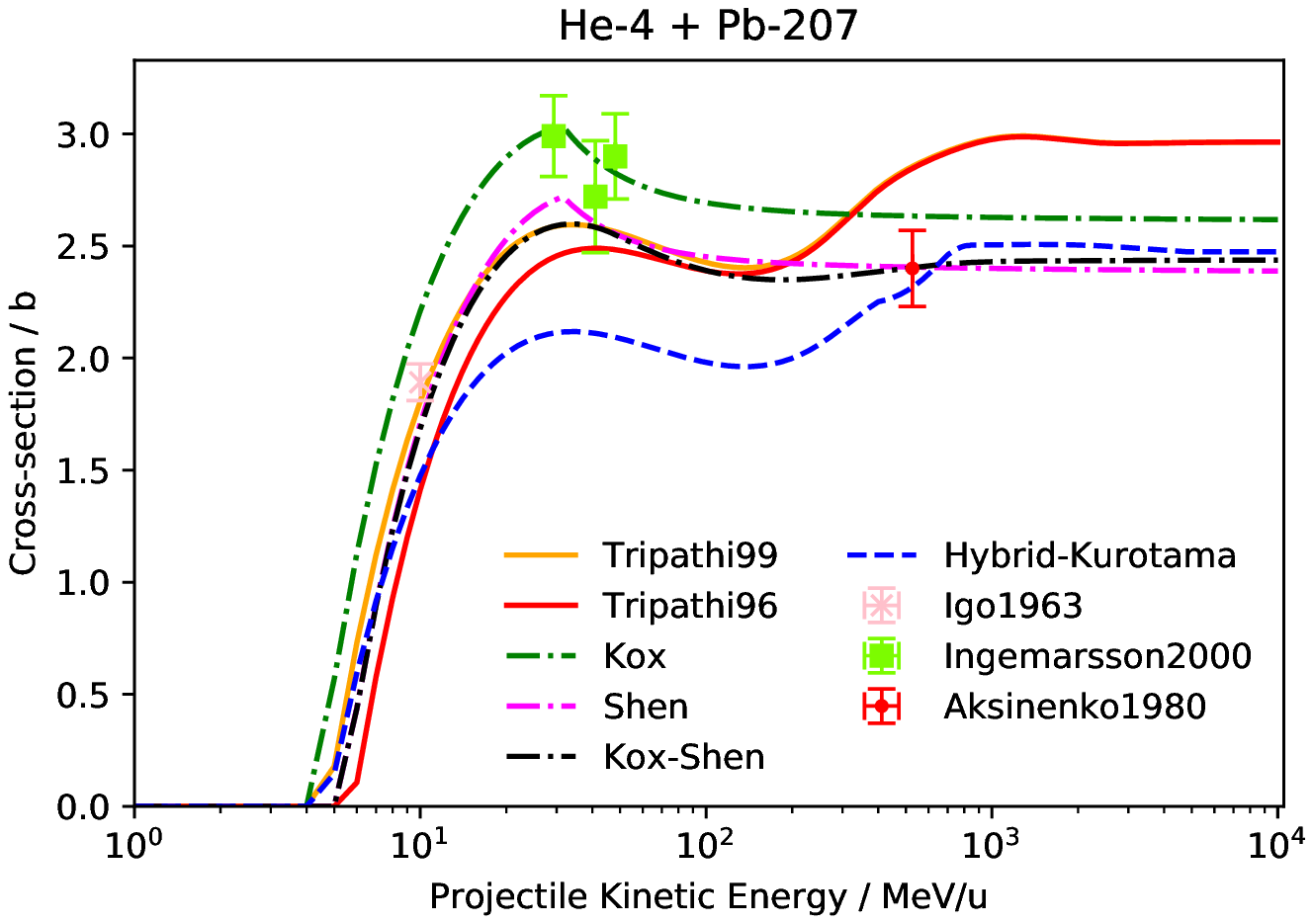}}
		\subfloat[][]
		{\includegraphics[width=.55\textwidth]{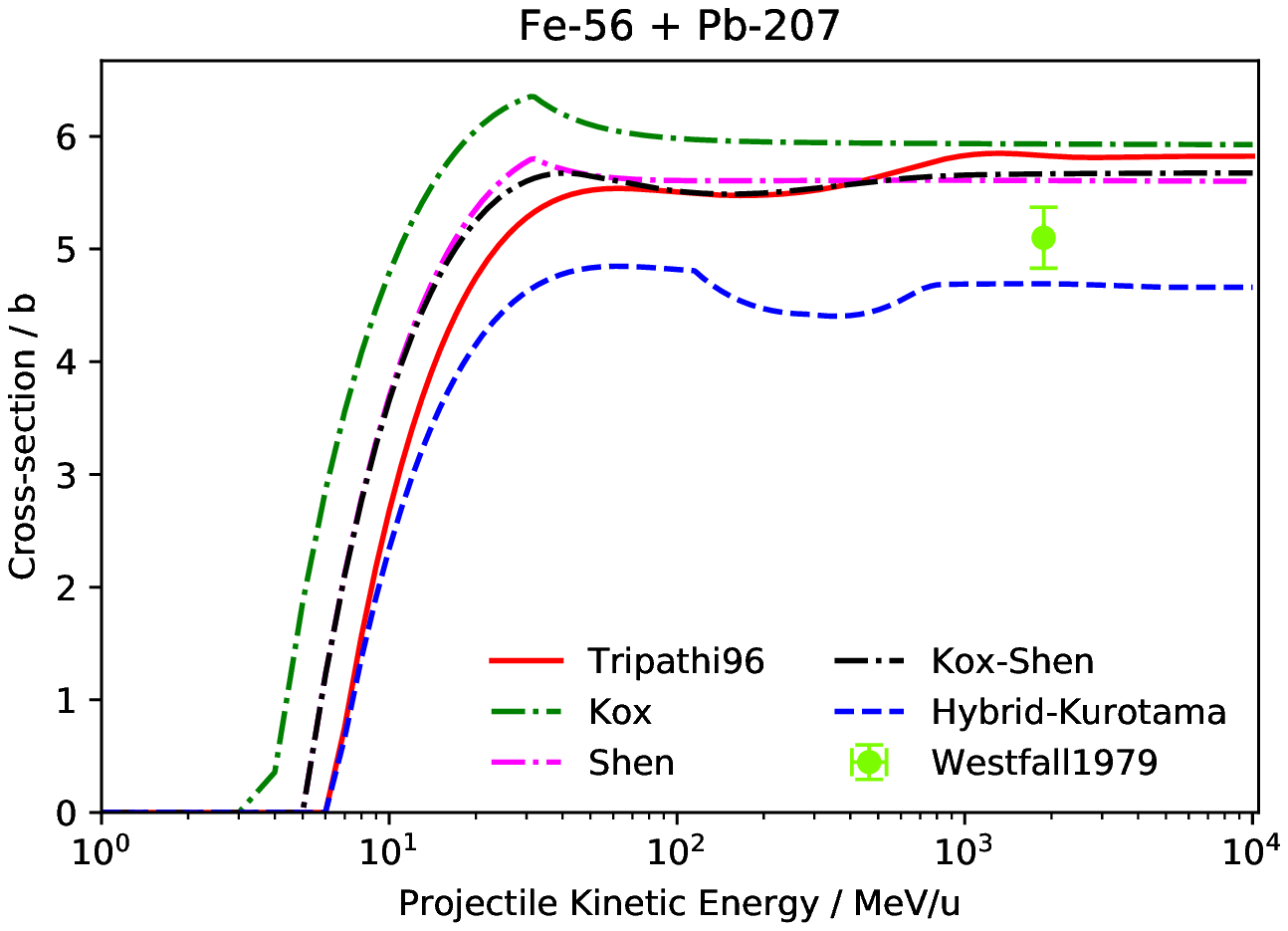}}
		\caption{Comparison between parameterization results and cross-section data for the following systems: $^{4}$He + $^{207}$Pb and $^{56}$Fe + $^{207}$Pb. Data in panel (a) are from references~\cite{Igo1963,Ingemarsson2000,Aksinenko1980}, in panel (b) from reference~\cite{Westfall1979}.}
		\label{fig:highE_data}
	\end{figure}

\begin{figure}[htb]
		\centering
		\includegraphics[width=0.7\textwidth]{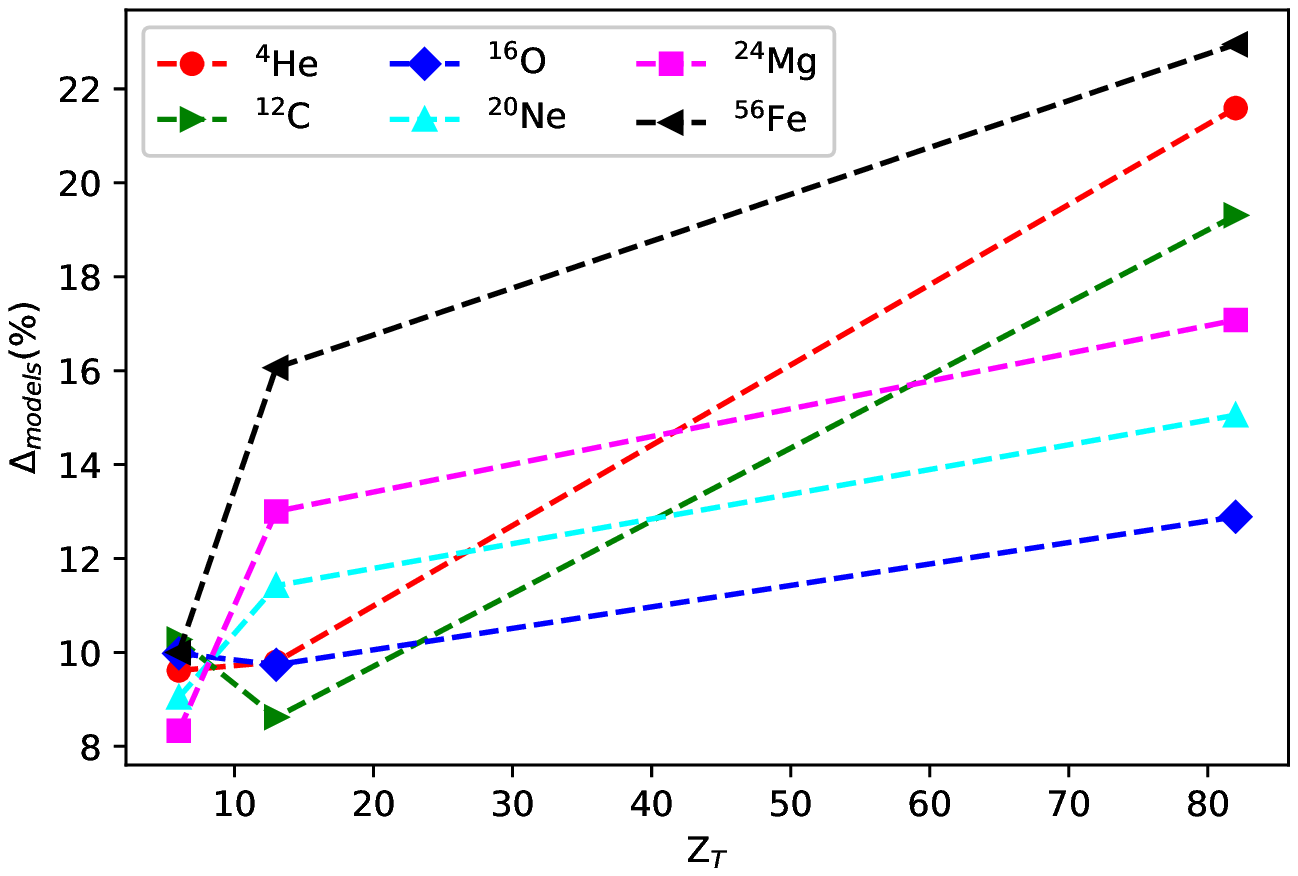}
		\caption{Relative differences between the maximum and minimum cross-section values predicted by the different models at $\SI{10}{GeV/u}$ (divided by the average value for all models) for different projectile nuclei as a function of the target atomic number. The dashed line connects the points to guide the reader's eyes.}
		\label{fig:highE_D}
	\end{figure}

	The models give different results at high energies (see Figure~\ref{fig:highE_data}). In particular, Tripathi96 tends to predict cross-section values that are higher than the results from the other models and for heavy projectiles Hybrid-Kurotama predicts values that are lower. The relative differences between the maximum and minimum cross-section values predicted by the different models at $\SI{10}{GeV/u}$ have been computed for several systems.  
	They are shown in Figure~\ref{fig:highE_D} for different projectile nuclei as a function of the target atomic number. The general trend is that such relative deviations are larger for heavier projectiles, and they become even larger the heavier the targets.
	Such differences underline uncertainties in the MC simulation results that come from the choice of the cross-section model. The heavier the nuclei under study, the larger the uncertainties. 
	
	\section{Discussion}
	\label{sec:recommendations}
	Human space exploration relies on MC simulation capabilities, which depend on the quality of the physics models implemented inside the MC codes.
	Thanks to the availability of the data collection to the community, cross-section parameterizations can be improved and, as a consequence, also simulation capabilities.
	
	\subsection{Parameterizations}
	All models are able to represent the trend of experimental data quite well. Only Tripathi96 $D$ value should be used without it being dividing by a factor $3$ for lithium projectiles.
	Nevertheless, Kox, Shen and Kox-Shen cannot reproduce a consistent drop in the datasets at intermediate energies. However, they can be easily implemented to work quite well without system-dependent fine tuning. The Kox model tends to overestimate the importance of the neutron excess parameter, which makes it more error prone when used for heavy-target systems.
	On the other hand, Tripathi96 tends to overestimate the data at high energies, Tripathi99 underestimates the data at low and intermediate energies, and the Hybrid-Kurotama model underestimates the intermediate energy values for $^4$He projectiles and for the $^{20}$Ne + $^{27}$Al system, and intermediate and high energy values for $^{56}$Fe projectiles.

	\subsection{Gaps in the experimental data}
	For several important systems in the field of radiation protection in space, either no cross-section data or not enough of them or non-reliable data are present in literature to be compared to the models at all energy ranges. Examples are $^{16}$O, $^{20}$Ne and $^{24}$Mg projectile data and the systems $^{56}$Fe + $^{16}$O, $^{27}$Al + $^{28}$Si, $^{28}$Si + $^{27}$Al and $^{28}$Si + $^{28}$Si. Since cross-section measurements on oxygen targets are difficult to perform, inverse kinematic measurements using $^{16}$O beams on various solid targets (e.g. carbon, aluminium, silicon, iron) could be very efficient to fill some of the gaps.
	High-energy ($>\SI{1}{GeV/u}$) data are available for almost none of the systems in figures~\ref{fig:He3} to~\ref{fig:Fe56}, meaning that the models cannot be validated at such energies. This is even more important considering the large differences between the predictions of different models at high energies (see Figure~\ref{fig:highE_D}). They are relevant when it comes to cosmic radiation transport through matter and can lead to large simulation errors. Therefore, it is important to measure high-energy cross-sections to improve the simulation capabilities.
	
	\section{Web application}
	\label{sec:webpage}
	A web application has been developed for the users to access the database. 
	Two tabs navigate the user to the related sections. The first section is a display of the data (``database'' section). In the second section, the user can directly plot data alongside the parameterizations described in section~\ref{sec:formulae} (``plot'' section). It also possible to plot data only.
	The sections are independent of each other. They are equipped with filters and settings. The filters allow the user to select the data and eventually, parameterizations of interest. Through the settings, the visual properties of the data collection and of any generated plots can be adjusted. 
	It is possible to download or send data and the generated plots to a desired email address.
	The plot section allows the user also to plot parameterizations only. In this case, the target can be any isotope or element or self-defined molecule. Atomic and mass numbers of such compound targets are computed as described in section~\ref{sec:DB}.
	The cross-sections of projectiles impinging on compound targets are computed as:
	\begin{equation}
		\sigma =  \sum_i n_i (\sigma_i),
		\label{Eq:CS_comp} 
	\end{equation}
	where $n_i$ is the number of atoms of the element $i$ in the molecule and $\sigma_i$ is the reaction cross-section of the projectile with the element. $\sigma_i$ is computed taking into account the natural isotopic abundances of the element $i$.
	
	
	The link to the web application is: \texttt{https://www.gsi.de/fragmentation} 
	
	\section{Conclusions}
	Total nucleus-nucleus reaction cross-sections are a key ingredient of deterministic and stochastic heavy-ion transport codes. Currently, there is no single parameterization able to reproduce all the measured data for all systems in all energy regions. In addition, not all datasets are so reliable.
	In the present work, a total nuclear reaction cross-section data collection for nucleus-nucleus systems is presented. All details about the implementation are given. The data collection can be found in the web application tool developed in the present work: \texttt{https://www.gsi.de/fragmentation}. Within the application, it is also possible to filter the data and generate plots.
	All reaction cross-sections extracted from the data collection have been compared to the different models for systems relevant to heavy-ion therapy and radiation protection in space applications.
	Generally speaking, all models tend to fit the experimental data well, but many datasets do not seem reliable for such a comparison with cross-section models. Kox, Shen and Kox-Shen are not able to represent the mid-energy ``valley'' that the data seem to show consistently. On the other hand, the Hybrid-Kurotama model underestimates the mid and high-energy values for heavy projectiles; Tripathi99 underestimates the data at low and mid energies; Tripathi96 tends to overestimate the data at high energies.
	Recommendations about what cross-section data should be measured to improve simulation capabilities are given in section~\ref{sec:recommendations}. Limitations of all currently available nuclear reaction cross-section parameterizations are the lack of projectile-target symmetry due to the neutron excess parameter 
	and the application of this parameter to all energy ranges for all models, where it is significant only for energies lower than \SI{200}{MeV/u} \cite{Kox1987}. 
	The lack of data in the high-energy ($>\SI{1}{GeV/u}$) region, where the models show significant differences, leads to large uncertainties in transport simulations.
	
	
	\section*{Acknowledgements}
	The authors acknowledge Dr. Maurizio Spurio, Dr. Z. Y. Sun, Dr. Jean-Éric Ducret, Dr. David Perez Loureiro, Dr. Jun-Sheng LI, Dr. Dong-Hai Zhang, Dr. Andrea Jungclaus, Dr. Ingrid Schall and Dr. Maya Takechi for providing further details about the data reported in their publications that have been included in the database.
	The authors also acknowledge Dr. Tatsuhiko Sato and Dr. Hiroshi Iwase for helping and supporting the implementation of Tripathi99 and Hybrid-Kurotama models, Dr. Albana Topi and Dr. Michael Kr\"amer for the help and suggestions regarding the data collection.
	The work was supported in the framework of the work package 200 of the ROSSINI3 project (ESA Contract No.4000125785/18/NL/GLC), which was a 2-year project started in December 2018, funded by ESA ESTEC and led by Thales Alenia Space Italia.
	
	\section*{References}
	\bibliography{References}
	\bibliographystyle{unsrt}

\end{document}